\documentclass[a4paper,12pt]{article}
\usepackage[utf8]{inputenc}
\usepackage[T1]{fontenc}
\usepackage{todonotes}
\usepackage{epsfig,graphicx,xcolor,amsbsy,amssymb,latexsym,amsfonts,amsmath,tcolorbox,setspace,mathrsfs}
\usepackage{pstricks}
\usepackage{enumerate}
\usepackage{color}
\usepackage{mathdots}
\usepackage{soul}
\usepackage{accents}
\usepackage{placeins}
\usepackage{wrapfig,framed,caption}
\usepackage[makeroom]{cancel}
\usepackage{amsfonts}
\usepackage{dsfont}
\usepackage{mmacells}
\mmaDefineMathReplacement{∞}{\infty}

\usepackage[style=numeric,sorting=none,style=numeric-comp,maxnames=10]{biblatex}

\definecolor{shadecolor}{gray}{0.925}

\usepackage{eurosym}
\usepackage{enumitem}
\usepackage[a4paper]{geometry}
\usepackage{graphicx}
\usepackage{tikz}
\usetikzlibrary{cd}
\usetikzlibrary{decorations.pathreplacing,decorations.markings,snakes}
\usetikzlibrary{decorations.pathmorphing}
\usepackage{xcolor}
\usepackage{amsmath,amssymb,amsfonts,pstricks,setspace}
\usepackage[enableskew]{youngtab}
\usepackage{ytableau}

\numberwithin{equation}{section}

\newcommand{\bea}{\begin{eqnarray}\displaystyle}
\newcommand{\eea}{\end{eqnarray}}

\newcommand{\figref}[1]{Fig.~\protect\ref{#1}}

\newcommand{\Qt}{Q_{\tau}}
\newcommand{\Qr}{Q_{\rho}}

\newenvironment{bsmallmatrix}
  {\left[\!\!\begin{smallmatrix}}
  {\end{smallmatrix}\!\!\right]}

\usepackage{float}
\newtcolorbox{summary}[2][]{colbacktitle=blue!10!white, colback=yellow!10!white,coltitle=blue!70!black, title={#2},fonttitle=\bfseries,#1}

\title{
{\bf Seiberg-Witten curves of $\widehat D$-type Little Strings}\\[40pt]}

\author{\large \textsc{Baptiste~Filoche\footnote{\tt b.filoche@ip2i.in2p3.fr}}~~,~~\textsc{Stefan~Hohenegger\footnote{\tt s.hohenegger@ipnl.in2p3.fr}}
~,~\,and\,~\textsc{Taro~Kimura\footnote{\tt taro.kimura@u-bourgogne.fr}}
}

\begin{document}

\maketitle
\thispagestyle{empty}
\begin{center}
\renewcommand{\thefootnote}{\fnsymbol{footnote}}\vspace{-0.5cm}
${}^{\footnotemark[1]\,\footnotemark[2]}$ Univ Lyon, Univ Claude Bernard Lyon 1, CNRS/IN2P3, IP2I Lyon, UMR 5822, F-69622, Villeurbanne, France\\[0.2cm]
${}^{\footnotemark[2]}$ Dept. of Physics E. Pancini, Universit\`a di Napoli Federico II, via Cintia, 80126 Napoli, Italy\\[0.2cm]
${}^{\footnotemark[3]}$ Institut de Math\'ematiques de Bourgogne, Universit\'e de Bourgogne, CNRS, UMR 5584, France\\[2.5cm]
\end{center}

\begin{abstract}
Little Strings are a type of non-gravitational quantum theories that contain extended degrees of freedom, but behave like ordinary Quantum Field Theories at low energies. A particular class of such theories in six dimensions is engineered as the world-volume theory of an M5-brane on a circle that probes a transverse orbifold geometry. Its low energy limit is a supersymmetric gauge theory that is described by a quiver in the shape of the Dynkin diagram of the affine extension of an ADE-group. While the so-called $\widehat{A}$-type Little String Theories (LSTs) are very well studied, much less is known about the $\widehat{D}$-type, where for example the Seiberg-Witten curve (SWC) is only known in the case of the $\widehat{D}_4$ theory. In this work, we provide a general construction of this curve for arbitrary $\widehat{D}_{M}$ that respects all symmetries and dualities of the LST and is compatible with lower-dimensional results in the literature. For $M=4$ our construction reproduces the same curve as previously obtained by other methods. The form in which we cast the SWC for generic $\widehat{D}_M$ allows to study the behaviour of the LST under modular transformations and provides insights into a dual formulation as a circular quiver gauge theory with nodes of $Sp(M-4)$ and~$SO(2M)$.
\end{abstract}

\newpage

\tableofcontents

\section{Introduction}
Little String Theories (LSTs) \cite{Witten:1995zh,Aspinwall:1997ye,Intriligator:1997dh,Hanany:1997gh,Brunner:1997gf} are a class of non-gravitational quantum theories, which at low energies behave like quantum field theories with point-like degrees of freedom, but whose UV-completion requires new extended degrees of freedom. Such theories arise naturally within the framework of String Theory, through suitable limits in which the gravitational sector is decoupled \cite{Aharony:1999ks,Kutasov:2001uf}. They constitute interesting generalisations of (conformal) field theories that are obtained through different point-particle limits and have therefore attracted a lot of attention over the years: for example, through geometric methods  similar to the case of (higher dimensional) conformal field theories, a classification has been provided in \cite{Bhardwaj:2015oru,Bhardwaj:2019hhd,Bhardwaj:2022ekc}.

A very rich class of supersymmetric LSTs (dubbed Little String Orbifolds in \cite{Hohenegger:2016eqy}) can be constructed as the world-volume theory of $N$ parallel M5-branes on a circle, probing a transverse orbifold geometry \cite{Blum:1997fw} (see also \cite{Haghighat:2013gba,Haghighat:2013tka,Hohenegger:2013ala,Hohenegger:2015cba,Hohenegger:2015btj,Hohenegger:2016yuv,Ahmed:2017hfr,Bastian:2017ing,Bastian:2017ary,Bastian:2017jje,Bastian:2018dfu,Bastian:2018fba,Bastian:2018jlf,Bastian:2019hpx,Bastian:2019wpx,Hohenegger:2019tii,Hohenegger:2020slq,Bastian:2019wpx,Hohenegger:2019tii,Hohenegger:2020gio,Filoche:2022qxk,Filoche:2023yfm}) of the form $\mathbb{R}_\perp^4/\Gamma$, where $\Gamma\subset SU(2)$. At low energies, the world-volume theory on the M5-branes is a six-dimensional supersymmetric gauge theory, described by a quiver in the shape of the Dynkin diagram of the affine group $\widehat{\Upsilon}$, which is related to $\Gamma$ through the McKay correspondance. Without additional flavour groups, the rank of the gauge nodes is fixed through the absence of anomalies and matter appears in the bifundamental representation.

Among these theories the case with $\widehat{\Upsilon}=\widehat{A}_{M-1}$ (called $\widehat{A}$-type and denoted $(\widehat{A}_{M-1},N)$ theory in the following) has been studied in quite some detail: indeed, this class of theories enjoys many different dual descriptions, which allow to calculate important quantities explicitly. For example, the BPS-(instanton) partition function can be obtained by counting M2-branes stretched between the M5 branes \cite{Haghighat:2013gba,Haghighat:2013tka,Hohenegger:2013ala,Hohenegger:2015cba}. Using a dual description in terms of F-theory compactified on a toric Calabi-Yau manifold $X_{N,M}$ \cite{Kanazawa:2016tnt}, this counting is captured by the topological string partition function on this threefold \cite{Hohenegger:2015btj,Hohenegger:2016eqy,Hohenegger:2016yuv,Bastian:2017ing}, which in turn can be computed in an algorithmic fashion using the (refined) topological vertex \cite{Aganagic:2003db,Iqbal:2007ii}. Another quantity of the $(\widehat{A}_{M-1},N)$ theory, which can be constructed explicitly using various different approaches \cite{Braden:2003gv,Hollowood:2003cv}, is the Seiberg-Witten curve (SWC). The latter encodes important information about the non-perturbative structure of the theory, notably its symmetries. Furthermore, it affords interesting insights into related algebraic integrable systems~\cite{Donagi:1995cf,Seiberg:1996nz,Koroteev:2019gqi}, where it is identified with their spectral curve. The SWC depends on $MN+2$ independent parameters and can be written as a linear combination of a basis of theta functions that are defined on a genus 2 surface. This surface reflects the doubly periodic nature of the M5-brane setup (which is for example also visible in the double elliptic fibration structure of the threefold $X_{N,M}$ mentioned above). Furthermore, for a generic $(\widehat{A}_{M-1},N)$ theory, this surface has a full period matrix: the third parameter, besides the two modular parameters, has an interpretation as the mass of the bifundamental matter from the perspective of the quiver gauge theory.

In this work, however, we shall be concerned with the case $\widehat{\Upsilon}=\widehat{D}_{M}$ (for $M\geq 4$) and $N=1$ (\emph{i.e.} a single M5-brane), which we shall denote as $(\widehat{D}_M,1)$ theory in the following. Due to technical difficulties and subtleties in realising dual descriptions, these so-called $\widehat{D}$-type LSTs are much less studied in the literature. Explicit computations of the instanton partition function have been performed in the topological vertex formalism by using so-called trivalent gluing techniques \cite{Hayashi:2017jze,Hayashi:2021pcj,Wei:2022hjx}, using the blow-up equations \cite{Kim:2019uqw,Kim:2023glm} or using dual realisations in term of two-dimensional $\mathcal{N}=(0,4)$ gauge theories~\cite{Kim:2017xan}. From the perspective of the SWC, to date only the curve in the case $M=4$ is known \cite{Haghighat:2018dwe}, either from the thermodynamic limit \cite{Nekrasov:2003rj} of the instanton partition function or by using the quantum states formalism \cite{Haghighat:2018gqf}. 

For general $M$, the geometric structure of the SWC is somewhat less restrictive than in the $(\widehat{A}_{M-1},1)$ case: due to the absence of a mass parameter~\cite{Haghighat:2017vch}, a suitable set of basis functions consists of products of genus 1 theta functions (with two different modular parameters), rather than genuine genus 2 theta functions as in the $\widehat{A}$-type case. In this paper, we present a general ansatz for the SWC of the $(\widehat{D}_M,1)$ theory that is manifestly compatible with all symmetries. To further restrict this ansatz, we use the fact that the $(\widehat{D}_M,1)$ theory is dual to a circular quiver gauge theory with one gauge node of type $Sp(M-4)$ and one node of type $SO(2M)$. This requires that the SWC can equivalently be re-written in a way that is compatible with the symmetries of this dual theory. We argue for low values of $M$ (notably $M=4$ and $M=5$) that consistency of these dual presentations imposes further conditions on the parameters and ultimately leaves only a few distinct forms of the curve (two in the case of $M=4$ and three in the case of $M=5$). By further considering a dimensional reduction (by assuming a natural behaviour of the remaining parameters in the decompactification limit) and comparing with known results in five-dimensions~\cite{Hayashi:2023boy}, we find further conditions on the parameters leaving a unique form of the SWC. In particular in the case of $M=4$, the latter agrees with known results in \cite{Haghighat:2018dwe,Haghighat:2018gqf}. By studying further examples up to $M=12$, we find distinctive patterns, which allow us to conjecture the general form of the $(\widehat{D}_M,1)$ SWC for generic $M\geq 4$. The latter could be identified with the spectral curves of new $\widehat D$-type double elliptic integrable systems. Some steps have been taken in this direction through the construction of the quantum integrable system of $D$-type conformal matter (\emph{i.e.} six-dimensional Super Conformal Field Theories (SCFTs)) \cite{Chen:2023aet} and quantum deformations of SWCs \cite{Chen:2020jla,Chen:2021ivd,Chen:2021rek}.

This paper is organised as follows: In Section~\ref{Sect:SWGeometry}, we provide more details on LSTs engineered from M5-branes probing transverse orbifold geometries and their SWCs, recalling in particular the $(\widehat{A}_{M-1},N)$ theories. In Section~\ref{sec:DSWC} we focus on the $(\widehat{D}_M,1)$ LSTs and present a general ansatz for the SWC that is compatible with all symmetries of the theory. In Section~\ref{sec:d4} we discuss in detail the case $M=4$ and show how this ansatz can be further restricted by imposing duality of the $(\widehat{D}_4,1)$ theory to a quiver gauge theory with one $Sp(0)$ and one $SO(8)$ node and by comparing to known five-dimensional results. We show that the unique SWC obtained in this way is equivalent to the result in \cite{Haghighat:2018dwe,Haghighat:2018gqf}. In Section~\ref{sec:d5} we repeat this construction for the $(\widehat{D}_5,1)$ theory and show that it again leads to a unique SWC, which constitutes a novel result in the literature. Based on further examples up to $M=12$, we present in Section~\ref{sec:generalisation} a general form for the SWC of the $(\widehat{D}_M,1)$ theory. Finally, Section~\ref{Sect:Conclusions} contains our conclusions, as well as an outlook on applications of our results. This paper is accompanied by two Appendices: in Appendix~\ref{App:ModularPropertie} we compile definitions as well as useful identities for modular objects that are used throughout the main body of this text. Appendix~\ref{app:Mmat} contains the results of the SWC for certain $M>5$, which corroborate our general conjecture in Section~\ref{sec:generalisation}.

\section{Little String Theories and Seiberg-Witten curves}\label{Sect:SWGeometry}

\subsection{$\widehat{ADE}$-type Little String Theories}\label{subsec:braneset}
Little String Theories (LSTs) arise in string theory (or related dual descriptions) through different decoupling limits \cite{Witten:1995zh,Aspinwall:1997ye,Seiberg:1997zk,Intriligator:1997dh,Hanany:1997gh,Brunner:1997gf} (see also \cite{Aharony:1999ks,Kutasov:2001uf} for reviews). They follow an $\widehat{ADE}$-classification \cite{Bhardwaj:2015oru} and can be engineered from numerous setups. For concreteness, we begin with a construction in M-theory \cite{Hohenegger:2015btj,Hohenegger:2016eqy,}, consisting of $N$ M5-branes (along directions $x^{0,1,\ldots,5}$), separated along a circle $\mathbb{S}_6^1$ probing a transverse $\mathbb R^4_\perp/\Gamma$ orbifold with a finite subgroup of $SU(2)$, $\Gamma \subset SU(2)$:
\begin{center}
    \begin{tabular}{c|cc|ccc|c|c|cccc}
         & $\mathbb S^1_0$ & $\mathbb S^1_1$ &  &$\mathbb R^3_{\parallel}$ & & $\mathbb S_5^1$ &$\mathbb S_6^1$ & &$\mathbb R^4_\perp \!\!\!$ &$\!\!\!\!\!\!/\Gamma$ &   \\[4pt]
         & $x^0$ & $x^1$ & $x^2$ &$x^3$ &$x^4$ &$x^5$ &$x^6$ &$x^7$ &$x^8$ &$x^9$ &$x^{10}$   \\\hline
        $N$ $M5$ & $=$ &$=$ &$=$ &$=$ &$=$ &$=$ &$\times$&&&&
    \end{tabular}
    \label{tab:m5setup}
\end{center}
In the low energy limit, this construction corresponds to a supersymmetric quiver gauge theory on $\mathbb{S}_0^1\times \mathbb{S}_1^1\times \mathbb R^3_\parallel\times \mathbb{S}_5^1$. The McKay correspondence \cite{Douglas:1996sw} associates the orbifold group $\Gamma$ to an affine quiver group $\widehat \Upsilon$ as follows

\begin{figure}[htbp]
\scalebox{0.6}{\parbox{5.5cm}{\begin{tikzpicture}
\draw [ultra thick,domain=0:200] plot ({2*cos(\x)}, {2*sin(\x)});
\draw [ultra thick,dashed,domain=200:250] plot ({2*cos(\x)}, {2*sin(\x)});
\draw [ultra thick,domain=250:360] plot ({2*cos(\x)}, {2*sin(\x)});
\draw[ultra thick,fill=white] (2,0) circle (0.5cm);
\node at (2,0) {$N$};
\draw[ultra thick,fill=white] ({2*cos(45)}, {2*sin(45)}) circle (0.5cm);
\node at ({2*cos(45)}, {2*sin(45)}) {$N$};
\draw[ultra thick,fill=white] ({2*cos(90)}, {2*sin(90)}) circle (0.5cm);
\node at ({2*cos(90)}, {2*sin(90)}) {$N$};
\draw[ultra thick,fill=white] ({2*cos(135)}, {2*sin(135)}) circle (0.5cm);
\node at ({2*cos(135)}, {2*sin(135)}) {$N$};
\draw[ultra thick,fill=white] ({2*cos(180)}, {2*sin(180)}) circle (0.5cm);
\node at ({2*cos(180)}, {2*sin(180)}) {$N$};
\draw[ultra thick,fill=white] ({2*cos(270)}, {2*sin(270)}) circle (0.5cm);
\node at ({2*cos(270)}, {2*sin(270)}) {$N$};
\draw[ultra thick,fill=white] ({2*cos(315)}, {2*sin(315)}) circle (0.5cm);
\node at ({2*cos(315)}, {2*sin(315)}) {$N$};
\node at (0,0) {\LARGE $\widehat A_M$};
\end{tikzpicture}
}}
\scalebox{0.6}{\parbox{5.5cm}{\begin{tikzpicture}
\draw [ultra thick] (0,1) -- (1,0);
\draw [ultra thick] (0,-1) -- (1,0);
\draw [ultra thick,dashed] (1,0) -- (3,0);
\draw [ultra thick] (4,1) -- (3,0);
\draw [ultra thick] (4,-1) -- (3,0);
\draw[ultra thick,fill=white] (1,0) circle (0.5cm);
\node at (1,0) {$2N$};
\draw[ultra thick,fill=white] (3,0) circle (0.5cm);
\node at (3,0) {$2N$};
\draw[ultra thick,fill=white] (0,1) circle (0.5cm);
\node at (0,1) {$N$};
\draw[ultra thick,fill=white] (0,-1) circle (0.5cm);
\node at (0,-1) {$N$};
\draw[ultra thick,fill=white] (4,1) circle (0.5cm);
\node at (4,1) {$N$};
\draw[ultra thick,fill=white] (4,-1) circle (0.5cm);
\node at (4,-1) {$N$};
\node at (2,1) {\LARGE $\widehat D_M$};
\end{tikzpicture}
}}
\scalebox{0.6}{\parbox{4.5cm}{\begin{tikzpicture}
\draw [ultra thick] (0,0) -- (0,6);
\draw [ultra thick] (0,3) -- (3,3);
\draw[ultra thick,fill=white] (0,0) circle (0.5cm);
\node at (0,0) {$N$};
\draw[ultra thick,fill=white] (0,1.5) circle (0.5cm);
\node at (0,1.5) {$2N$};
\draw[ultra thick,fill=white] (0,3) circle (0.5cm);
\node at (0,3) {$3N$};
\draw[ultra thick,fill=white] (0,4.5) circle (0.5cm);
\node at (0,4.5) {$2N$};
\draw[ultra thick,fill=white] (0,6) circle (0.5cm);
\node at (0,6) {$N$};
\draw[ultra thick,fill=white] (1.5,3) circle (0.5cm);
\node at (1.5,3) {$2N$};
\draw[ultra thick,fill=white] (3,3) circle (0.5cm);
\node at (3,3) {$N$};
\node at (1.5,4.3) {\LARGE $\widehat E_6$};
\end{tikzpicture}
}}
\scalebox{0.6}{\parbox{5.5cm}{\begin{tikzpicture}
\draw [ultra thick] (0,0) -- (9,0);
\draw [ultra thick] (4.5,0) -- (4.5,1.5);
\draw[ultra thick,fill=white] (0,0) circle (0.5cm);
\node at (0,0) {$N$};
\draw[ultra thick,fill=white] (1.5,0) circle (0.5cm);
\node at (1.5,0) {$2N$};
\draw[ultra thick,fill=white] (3,0) circle (0.5cm);
\node at (3,0) {$3N$};
\draw[ultra thick,fill=white] (4.5,0) circle (0.5cm);
\node at (4.5,0) {$4N$};
\draw[ultra thick,fill=white] (6,0) circle (0.5cm);
\node at (6,0) {$3N$};
\draw[ultra thick,fill=white] (7.5,0) circle (0.5cm);
\node at (7.5,0) {$2N$};
\draw[ultra thick,fill=white] (9,0) circle (0.5cm);
\node at (9,0) {$N$};
\draw[ultra thick,fill=white] (4.5,1.5) circle (0.5cm);
\node at (4.5,1.5) {$2N$};
\node at (4.5,-1.2) {\LARGE $\widehat E_7$};
\draw [ultra thick] (0,3) -- (10.5,3);
\draw [ultra thick] (3,3) -- (3,4.5);
\draw[ultra thick,fill=white] (0,3) circle (0.5cm);
\node at (0,3) {$2N$};
\draw[ultra thick,fill=white] (1.5,3) circle (0.5cm);
\node at (1.5,3) {$4N$};
\draw[ultra thick,fill=white] (3,3) circle (0.5cm);
\node at (3,3) {$5N$};
\draw[ultra thick,fill=white] (4.5,3) circle (0.5cm);
\node at (4.5,3) {$5N$};
\draw[ultra thick,fill=white] (6,3) circle (0.5cm);
\node at (6,3) {$4N$};
\draw[ultra thick,fill=white] (7.5,3) circle (0.5cm);
\node at (7.5,3) {$3N$};
\draw[ultra thick,fill=white] (9,3) circle (0.5cm);
\node at (9,3) {$2N$};
\draw[ultra thick,fill=white] (10.5,3) circle (0.5cm);
\node at (10.5,3) {$N$};
\draw[ultra thick,fill=white] (3,4.5) circle (0.5cm);
\node at (3,4.5) {$3N$};
\node at (4.5,4.3) {\LARGE $\widehat E_8$};
\end{tikzpicture}
}}
\caption{\sl $N$-coloring of $\widehat A$, $\widehat D$ and $\widehat E$-type quivers. }
\label{Fig:QuiverColoring}
\end{figure}
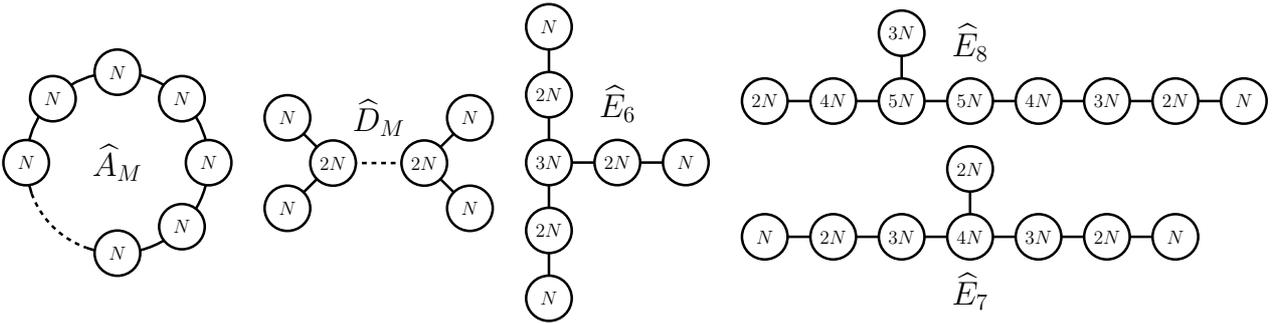

\begin{center}
    \begin{tabular}{c||c|c|c|c|c}
    Orbifold group $\Gamma$ & $\mathbb Z_M$ for $M\geq 1$ & $\mathbb{BD}_{M-2}$ for $M\geq 4$ & $\mathbb{BT}$ & $\mathbb{BO}$ & $\mathbb{BI}$\\ \hline
    &&&&&\\[-12pt]
    Quiver group $\widehat \Upsilon$ & $\widehat A_{M-1}$ & $\widehat D_{M}$ & $\widehat E_6$ & $\widehat E_7$ & $\widehat E_8$
    \end{tabular}
    \label{tab:McKay}
\end{center}
where $\mathbb{BD}_{M}$ is the binary dihedral group of order $4M$ and $\mathbb{BT},\mathbb{BO}$ and $\mathbb{BI}$ are respectively the binary isomorphism groups of the tetrahedron, the cube and the dodecahedron. The quiver takes the form of the Dynkin diagram of $\widehat{\Upsilon}$ (see~\figref{Fig:QuiverColoring}) and mathematically is described by $(\Upsilon_0,\Upsilon_1)$, where $\Upsilon_0$ is the set of nodes and $\Upsilon_1$ the set of edges connecting the latter. \figref{Fig:QuiverColoring} also display an $n$\emph{-coloring} of the quivers, \emph{i.e.} a particular assignment of the rank of each gauge node. To explain this, we introduce the {\it Cartan matrix} of the quiver $(\Upsilon_0,\Upsilon_1)$ by
\begin{equation}\label{eq:CartanMatrix}
    c_{ij} = 2\delta_{ij} - \sum_{e\in \Upsilon_1 :i \to j} 1 - \sum_{e\in \Upsilon_1 :j \to i} 1 \,, \quad\quad \forall i,j \in \Upsilon_0\,.
\end{equation}
The rank of the gauge group at each node in $\Upsilon_0$ is uniquely fixed\footnote{Here we do not assume any additional flavor groups.} by demanding the vanishing of the beta-functions for the gauge couplings  \cite{Lawrence:1998ja}:
\begin{equation}\label{eq:anomalyfree}
    \sum_{j \in \Upsilon_0} c_{ij} \cdot n_j = 0 \,, \quad \quad \forall i \in \Upsilon_0 \, ,
\end{equation}
with $n_j$ the rank of gauge group associated with the $j$-th node. A solution of the set of equations~\eqref{eq:anomalyfree} is called an $  n$-coloring of the quiver. For quivers of simply laced Lie groups, the coloring is fixed by a single integer $N$ which corresponds to the number of M5-branes probing the orbifold. Concretely, in~\figref{Fig:QuiverColoring}, the nodes \scalebox{0.4}{\parbox{1cm}{
\begin{tikzpicture} 
\draw[ultra thick,fill=white] (10.5,3) circle (0.5cm);
\node at (10.5,3) {$N$};
\end{tikzpicture}
}} of the quiver represent gauge groups $\widehat A_{N-1}$, while lines denote matter in the bifundamental representation.\footnote{Certain quiver gauge theories, such as the $\widehat{A}_0$ theory with $M=1$ has matter in the adjoint- rather than the bifundamental representation \cite{Hohenegger:2015btj}. However, we shall not consider this case in the current paper.} We shall denote a LST in this framework as $(\widehat\Upsilon, N)$. 

\subsection{$\widehat{A}$-type Seiberg-Witten curves}
Reviewing first the case of $\Gamma=\mathbb{Z}_M$, the so-called A-type LSTs engineered from the M-brane setup in Table~\ref{tab:m5setup} can be described by a doubly elliptic Seiberg-Witten curve \cite{Braden:2003gv,Hollowood:2003cv}. To understand the origin of the latter, we recall \cite{Haghighat:2013gba,Haghighat:2013tka,Hohenegger:2013ala,Hohenegger:2015cba,Hohenegger:2015btj,Hohenegger:2016eqy} that this M-brane configuration can be described by $MN+2$ independent parameters: we refer the reader for more details to \cite{Hohenegger:2015btj,Hohenegger:2016eqy} and only mention that in the following we denote the (complexified) radii of $\mathbb{S}_1^1$ and $\mathbb{S}_6^1$ (measured in units of the radius of $\mathbb{S}_0^1$) by $\tau$ and $\rho$ respectively. Furthermore, following \cite{Hohenegger:2015btj}, we introduce a $U(1)_S$ deformation of $\mathbb{R}_\perp^4/\mathbb{Z}_M$ with respect to $\mathbb{S}_0^1$
\begin{align}
U(1)_S:\hspace{0.4cm}(w_+,w_-)\longrightarrow\left(e^{2\pi i S}\,w_+\,,e^{-2\pi i S}\,w_-\right)\,,&&\text{with} &&w_\pm =x^7\pm i x^8\,,
\end{align}
which is compatible with the $\mathbb{Z}_M$ action. The parameters $(\tau,\rho,S)$ can be arranged into the period matrix $\Omega$ of a genus 2 Riemann surface and the SWC (parametrised by $(z_1,z_2)$) can be written in terms of genus $2$ theta functions $\Theta$ (see (\ref{eq:THETA}) for the definition) \cite{Braden:2003gv,Hollowood:2003cv}:
\begin{equation}\label{eq:genus2AA}
    \sum_{i=0}^{M-1} \sum_{j=0}^{N-1} a_{ij} \,\Theta \! \begin{bmatrix}
        i/M & j/N\\ 0 & 0
    \end{bmatrix}\! ( z \, | \Omega) =0\,,\quad z=\begin{bmatrix}z_1 \\ z_2\end{bmatrix},\quad \Omega = \begin{bmatrix}
        \frac{\tau}{M} & S \\ S & \frac{\rho}{N} 
    \end{bmatrix}.
\end{equation}
The $\{a_{ij}\}$ in (\ref{eq:genus2AA}) depend on the remaining parameters of the M-brane configuration mentioned above.\footnote{Indeed, due to an overall scaling symmetry of~\eqref{eq:genus2AA}, the moduli space of the SWC is indeed $(NM+2)$-dimensional. } Using a dual description in terms of F-theory compactified on a Calabi-Yau threefold $X_{N,M}$ \cite{Hohenegger:2015btj,Hohenegger:2016yuv,Hohenegger:2016eqy} with double elliptic-fibration structure, the $\{a_{ij}\}$ have been characterised in~\cite{Kanazawa:2016tnt} as (the product of) generating functions of Gromov-Witten invariants.

From the perspective of the $\widehat{A}_M$ quiver gauge theory represented by the leftmost diagram in \figref{Fig:QuiverColoring}, the parameter $\tau$ is interpreted as a complexified gauge coupling (which is common to all $M$ nodes in the quiver), $\rho$ as the (common) parameter which extends all gauge group nodes to their affine forms $\widehat{A}_{N-1}$ and $S$ as the mass of the bifundamental hypermultiplets. The remaining parameters \cite{Kanazawa:2016tnt} (appearing in the form of $a_{ij}$ in (\ref{eq:genus2AA})) encode further structure of the gauge- and matter spectrum, but the details shall not be important in the following. At the classical level, the theory inherits modular properties from the genus $2$ formulation~\eqref{eq:genus2AA}. The functional basis (\emph{i.e.} the particular set $\Theta \! \left[\begin{smallmatrix}
        i/M & j/N\\ 0 & 0
    \end{smallmatrix}\!\right]\! ( z \, | \Omega)$ of genus $2$ theta functions)  over which the curve \eqref{eq:genus2AA} is written, corresponds to theta functions of polarised Abelian varieties~\cite{book:Igusa,birkenhake2004complex}. The modular group acting on the period matrix $\Omega$ is therefore the paramodular group defined as \cite{birkenhake2004complex}:
\begin{equation}\label{eq:paramodularMN}
    \Sigma_{M,N}:= Sp(4,\mathbb Q) \cap \begin{bmatrix}
        \mathbb Z & \mathbb Z & \mathbb Z/M & \mathbb Z/N\\
        \mathbb Z & \mathbb Z & \mathbb Z/M & \mathbb Z/N\\
        M\mathbb Z & M\mathbb Z & \mathbb Z & M\mathbb Z/N\\
        N\mathbb Z & N\mathbb Z & N\mathbb Z/M & \mathbb Z
    \end{bmatrix}\,,
\end{equation}
with,
\begin{equation}
    Sp(4,\mathbb Q) = \left\{ \Lambda = \begin{bmatrix}
        A & B \\ C & D
    \end{bmatrix} \in \mathbb M_{4}(\mathbb Q)\,| \, \Lambda \begin{bmatrix}
        0 & \mathds{1}\\ \mathds{-1} & 0
    \end{bmatrix} \Lambda^T = \begin{bmatrix}
        0 & \mathds{1}\\ \mathds{-1} & 0
    \end{bmatrix} \right\},
\end{equation}
where $ \mathbb M_{4}(\mathbb Q)$ is the space of $4\times 4$ matrices with rational entries and non-vanishing determinant. The group $\Sigma_{M,N}$ acts in the following way on the period matrix $\Omega$:
\begin{equation}\label{eq:paramodularaction}
    \Omega \to (A\Omega +B)(C \Omega + D)^{-1}\,, \quad \forall \begin{bmatrix}
        A & B \\ C & D
    \end{bmatrix} \in \Sigma_{M,N}\,.
\end{equation}
It is interesting to note that the linearly acting subgroup of the paramodular group (\emph{i.e.} the subgroup with $B=C=0$ in~\eqref{eq:paramodularaction}) generates the web of dualities in LSTs that was described at the level of the instanton partition function in~\cite{Hohenegger:2016yuv,Bastian:2017ary,Bastian:2017ing}. In this respect, the group $\Sigma_{M,N}$ is a generalisation of the paramodular group found in \cite{Bastian:2019hpx,Bastian:2019wpx} for generic $M,N>1$. 

The $(\widehat A_{M-1},N)$ theory was shown to be dual to $(\widehat A_{M^\prime-1},N')$ for any $N',M'\in\mathbb{N}$ such that $N^\prime M^\prime = NM$ and $\mathrm{gcd}(N^\prime,M^\prime) = \mathrm{gcd}(N,M)=:p$. The corresponding duality map at the level of the period matrix $\Omega$, was formulated in \cite{Hohenegger:2016yuv} in terms of the following $Sp(4,\mathbb{Z})$ transformation
\begin{equation}\label{eq:mapNMNM}
    A = \begin{bmatrix}
        \frac{M}{p} & -\frac{N}{p} \\ 1 & -1
    \end{bmatrix}\,, \quad \Omega \to (A^T)^{-1} \Omega A^{-1} = \begin{bmatrix}
        \frac{\tau^\prime}{M^\prime} & S^\prime \\ S^\prime & \frac{\rho^\prime}{N^\prime}
    \end{bmatrix}\,.
\end{equation}
This transformation acts naturally on the basis of genus 2 theta functions $\Theta \! \left[\begin{smallmatrix}
        i/M & j/N\\ 0 & 0
    \end{smallmatrix}\!\right]\! ( z \, | \Omega)$ (for $i \in \{0\ldots M-1\}$ and $j \in \{0\ldots N-1$) over which (\ref{eq:genus2AA}) is formulated: using~\eqref{eq:GAactiongenus2} and noticing the equivalence of the following sets

\begin{align}
    \Bigg\{ A\cdot \left(\frac{i}{M},\frac{j}{N}\right),& \, 0\leq i \leq M-1\,,\,0\leq j \leq N-1 \Bigg\} \nonumber\\
    &\cong \Bigg\{\left(\frac{k}{p}\!\!\!\!\mod 1,\frac{lp}{MN}\!\!\!\!\mod 1\right), \, 0 \leq k \leq p-1\,, \, 0 \leq l \leq \frac{MN}{p}-1\Bigg\}\,,
\end{align}
it indeed follows that~\eqref{eq:mapNMNM} (along with a suitable duality map of the remaining parameters $\{a_{ij}\}$) leaves the SWC (\ref{eq:genus2AA}) invariant.

\section{$\widehat{D}$-type Seiberg-Witten curves}\label{sec:DSWC}

In this Section, we review some of the key properties of $(\widehat D_M,1)$ LSTs, which allow us to formulate a general ansatz for their SWCs. In the following Sections~\ref{sec:d4} and \ref{sec:d5} we shall use this ansatz to provide the unique form of the curves in the cases $M=4$ and $M=5$ respectively, before generalising our results in Section~\ref{sec:generalisation} to generic $M\geq 4$.

\subsection{$\widehat D$-type LSTs}
\begin{figure}[htbp]
    \centering
    \scalebox{0.9}{\parbox{9cm}{\begin{tikzpicture}
        \draw [ultra thick] (0,1) -- (1,0);
        \draw [ultra thick] (0,-1) -- (1,0);
        \draw [ultra thick] (1,0) -- (3.25,0);
        \draw [ultra thick] (4.25,0) -- (5,0);
        \draw [ultra thick,dashed] (5,0) -- (2.5,0);
        \draw [ultra thick] (6,1) -- (5,0);
        \draw [ultra thick] (6,-1) -- (5,0);
        \draw[ultra thick,fill=white] (1,0) circle (0.55cm);
        \node at (1,0) {\footnotesize $SU(2)$};
        \draw[ultra thick,fill=white] (2.5,0) circle (0.55cm);
        \node at (2.5,0) {\footnotesize $SU(2)$};
        \draw[ultra thick,fill=white] (5,0) circle (0.55cm);
        \node at (5,0) {\footnotesize $SU(2)$};
        \draw[ultra thick,fill=white] (0,1) circle (0.55cm);
        \node at (0,1) {\footnotesize $SU(1)$};
        \draw[ultra thick,fill=white] (0,-1) circle (0.55cm);
        \node at (0,-1) {\footnotesize $SU(1)$};
        \draw[ultra thick,fill=white] (6,1) circle (0.55cm);
        \node at (6,1) {\footnotesize $SU(1)$};
        \draw[ultra thick,fill=white] (6,-1) circle (0.55cm);
        \node at (6,-1) {\footnotesize $SU(1)$};
        \end{tikzpicture}
        }}
        \scalebox{0.8}{\parbox{6cm}{\begin{tikzpicture}
        \draw[ultra thick,fill=white] (0,0) circle (2cm);
        \draw[ultra thick,fill=white] (-2,0) circle (0.9cm);
        \node at (-2,0) {\footnotesize $Sp(M-4)$};
        \draw[ultra thick,fill=white] (2,0) circle (0.9cm);
        \node at (2,0) {\small $SO(2M)$};
        \end{tikzpicture}
        }}
    \caption{Quiver diagram of $(\widehat D_M ,1)$ LST with $M-3$ $SU(2)$ nodes (which are indicated explicitly) and its dual $\widehat A_1$ base $Sp(M-4)-SO(2M)$ quiver theory}
    \label{fig:DMAN}
\end{figure}
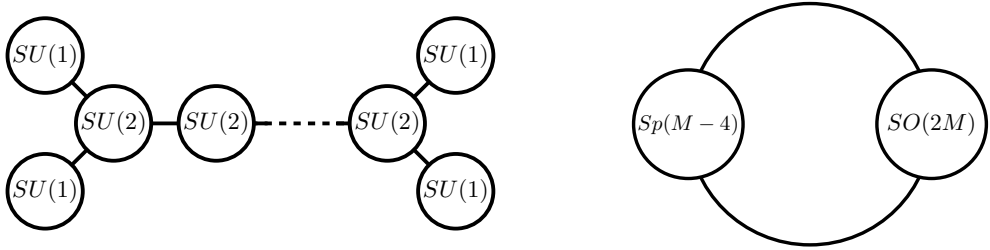

First of all, contrary to $\widehat A$-type, $\widehat D$-type LSTs do not allow for a  mass-deformation parameter \cite{Haghighat:2017vch}, \emph{i.e.} there is no equivalent to the parameter $S$ in the period matrix in (\ref{eq:genus2AA}). Therefore, also the genus 2 theta functions $\Theta$ are effectively reduced to two types of genus $1$ (\emph{i.e.} Jacobi) theta functions. Secondly, the transverse $\widehat D$-type orbifold itself has two important effects, which are of importance to the SWC: 
\begin{itemize}
\item[\emph{(i)}] the orbifold gives rise to a $\mathbb Z_2$ symmetry, under which the SWC is invariant \cite{Haghighat:2018dwe} 
\item[\emph{(ii)} ]{\it $M5$-brane fractionalisation}: as explained in \cite{DelZotto:2014hpa} due to the orbifold, a single $M5$-brane splits into two `half-branes' on the $\mathbb S_6^1$ circle of Table~\ref{tab:m5setup}. At the level of the quantum states introduced in \cite{Haghighat:2018gqf}, each $M5$ state can be described by the product of two $1/2$ $M5$-brane states.
\end{itemize}
Finally, there is another point of view, which provides crucial information for the SWCs, namely a dual description of the ($\widehat{D}_M,1)$ LSTs in terms of an $Sp(M-4)$--$SO(2M)$ quiver gauge theory, which is schematically shown in Figure~\ref{fig:DMAN}. Indeed, for $M=4$ this duality was first argued in \cite{Kim:2017xan} at the level of the partition function and in \cite{Haghighat:2018dwe} at the level of the SWC using fiber-base duality, while for general $M\geq 4$, it was established at the level of the partition function in \cite{Kim:2017xan}.

To understand the implications of such a duality at the level of the SWC, we remark that, in the absence of a mass-deformation, curves with $\widehat A_1$ base can generally be written in the following form \cite{Haghighat:2018dwe,Haghighat:2018gqf}:
\begin{equation}\label{eq:A1SOSP}
    0 = \theta_3^2(0;\rho) \theta_2^2(0;\rho) \theta_1^2(z_2;\rho) c_1(\tau)  F_1(z_1;\tau) - \theta_4^2(z_2;\rho)F_2(z_1;\tau),
\end{equation}
where the Jacobi theta functions $\theta_{2,3}$ are defined in \eqref{eq:defthetajac}, $c_1(\tau)$ is a modular constant and $F_{1,2}$ correspond to elliptic gauge polynomials\footnote{The gauge polynomial is the polynomial whose roots are the Coulomb branch moduli. Elliptic gauge polynomials are their natural elliptic uplift.} of each of the two nodes of the (right) quiver in Figure~\ref{fig:DMAN}. For example, in \cite{Haghighat:2018dwe}, these polynomials have been found for $SO(8)$ and $Sp(0)$ nodes to be respectively
\begin{align}
&F_1(z_1;\tau) = \prod_{i=1}^4 \frac{\theta_1(z_1 - a_i^{SO};\tau)\theta_1(z_1+a_i^{SO};\tau)}{\theta_1^2(z_1;\tau)}\,,&&\text{and} &&F_2(z_1;\tau) = \frac{\theta_1^2(2z_1;\tau)}{\theta_1^8(z_1;\tau)}\,.\label{eq:sso2m}
\end{align}

\noindent
The existence of the dual description of the $(\widehat{D}_M,1)$ theory requires that their SWC need to admit a presentation of the form (\ref{eq:A1SOSP}). As we shall see in the following, by starting from a generic ansatz (based on a finite dimensional basis of theta-function building blocks), this requirement leads to non-trivial constraints.

\subsection{General Ansatz for the Seiberg Witten Curve}\label{subsec:genansatz}
We are now ready to discuss the general form of the SWC of the $(\widehat D_M,1)$-LST. This curve is doubly elliptic and we shall denote the corresponding parameters by $\tau$ and $\rho$, while the coordinates shall be called $z_1$ (the coordinate on the $\tau$-parameterised elliptic curve) and $\mathbb Z_2$ even in $z_2$ (the coordinate on the $\rho$-parameterised elliptic curve). Following the discussion of the previous Subsection, we shall now provide a general ansatz for the $(\widehat D_M,1)$-LST, which is based on the following principles:
\begin{enumerate}
    \item The SWC is independently $\mathbb Z_2$ even in $z_1$ and $z_2$. This symmetry is due to the $\mathbb{Z}_2$ symmetry of the $\widehat{D}$-type orbifold, explained in point \emph{(i)} of the previous Subsection. The fact that it (a priori) can act independently on $z_1$ and $z_2$ is a consequence of the absence of a mass deformation in the $\widehat{D}$-type LSTs.
\item Using (pairs of) genus 1 theta functions as building blocks, the general form of the $\widehat{D}$-type-LST SWC can be cast in the following form
    \begin{equation}\label{eq:TMPHI}
        \sum_{0\leq i,j\leq M} \widetilde T_i(z_2,\rho,\tau)\, \mathbb{M}_{ij}(\tau)\, \Phi_j(z_1,\tau) = 0\,.
    \end{equation}
Indeed, here $\{\Phi_j\}_{j\in \{0\ldots M \}}$ is a basis of $\mathbb Z_2$ even theta functions of degree $2M$ (defined in~\eqref{eq:defPHI}) and $\widetilde T_i$ are $\mathbb Z_2$ even combinations of elliptic Weyl characters of $SU(1)$ and $SU(2)$ \cite{Haghighat:2017vch,Nekrasov:2012xe}, concretely
    \begin{equation}\label{eq:defTi}
        \widetilde T_i(z_2,\rho) = \begin{cases}
    \alpha_i \theta_1^2(z_2;\rho),& \text{for } i\in \{0,1,M-1,M\}\,,\\
    \alpha_i \theta_1^2(z_2;\rho) + \beta_i \theta_4^2 (z_2;\rho),& \text{for } i\in \{2\ldots M-2\}\,,
        \end{cases}
    \end{equation}
     with $\theta_{1}$ and $\theta_4$ defined eq.~\eqref{eq:defthetajac}. $\alpha_i, \beta_i$ are $2M-2=h^\vee(D_M)$, which are independent of $z_{1,2}$, but may depend on $\tau$ and $\rho$ (as well as other Coulomb moduli of the $\widehat{D}_M$-LST).
   
    \item The form (\ref{eq:TMPHI}) of the SWC must permit an alternative decomposition either on sections of line bundles with polarisation $(2,2M)$ or on products of sections of line bundles with polarisation $(1,M)$. As detailed in \cite{Haghighat:2018gqf}, this is interpreted in term of quantum states as a consequence of the $M5$-brane fractionalisation discussed in point \emph{(ii)} of the previous Subsection. This can alternatively be motivated by analysing the 4d SWC of the nodes involved in the dual construction (corresponding to the quiver on the right of~\figref{fig:DMAN}). The gauge polynomial of $SO(2M)$ and $Sp(M-4)$ gauge theories can be decomposed into the product of two degree $M$ polynomials \cite{DHoker:1996kdj}. We assume that this structure stems from a decomposition in term of sections of line bundle as explained above in the SWC of the LST.
\end{enumerate}

\noindent
The most general form of the SWC that is compatible with points 1. and 3. and permits a presentation of the form (\ref{eq:TMPHI}) in point 2., with the correct structure for the (elliptic) Weyl characters $\widetilde T_i$ takes the following form
\begin{equation}\label{eq:swcdmansatz}
\begin{split}
    \theta_1^2(z_2;\rho) &\left[ \sum_{0\leq i \leq j \leq \lfloor \frac{M}{2} \rfloor} a_{ij}^{(+)} X_i(z_1;\tau)X_j(z_1;\tau) +  \sum_{1\leq i \leq j \leq \lfloor \frac{M-1}{2} \rfloor} a_{ij}^{(-)} Y_i(z_1;\tau)Y_j(z_1;\tau)\right]\\
    &+\theta_4^2(z_2;\rho)\left[ \sum_{0\leq i \leq j \leq \lfloor \frac{M}{2} \rfloor} b_{ij}^{(+)} X_i(z_1;\tau)X_j(z_1;\tau) +  \sum_{1\leq i \leq j \leq \lfloor \frac{M-1}{2} \rfloor} b_{ij}^{(-)} Y_i(z_1;\tau)Y_j(z_1;\tau)\right]=0\,,
\end{split}
\end{equation}
where $X_i$ and $Y_i$ (with $i\in\{0,\ldots,\lfloor \frac{M-1}{2} \rfloor$) are defined in (\ref{eq:defxiyi}) in Appendix~\ref{Sect:ThetaFunctions}. Here $a_{ij}^{\pm}$ and $b_{ij}^{\pm}$ are a total of $(M+1)^2/2 + 3/2$ parameters for $M$ even and $(M+1)^2/2$ parameters for $M$ odd, which exceeds the expected number of $2M-2$. This implies that these parameters are not all independent and the ansatz (\ref{eq:swcdmansatz}) is too general. We therefore require further conditions to uniquely determine the SWC: such conditions are provided  by demanding an alternative presentation of the curve (\ref{eq:swcdmansatz}) in the form of (\ref{eq:A1SOSP}), compatible with a dual description as a quiver theory such as in the right part of Figure~\ref{fig:DMAN}. Finally, any remaining ambiguities can be fixed by comparison with known results of the 5-dimensional compactification of the LSTs. To explain this reasoning concretely, we shall explain in detail in the following Section~\ref{sec:d4} the case of the $(\widehat{D}_4,1)$ theory (for which the SWC was previously constructed in \cite{Haghighat:2018dwe,Haghighat:2018gqf}) and in Section~\ref{sec:d5} the case of the $(\widehat{D}_5,1)$ theory (for which the SWC is not known in the literature). Finally, in Section~\ref{sec:generalisation}, we present the generalisation of our results to theories $(\widehat{D}_M,1)$, with generic $M\geq 4$.

\section{Seiberg-Witten Curve of $(\widehat D_4, 1)$}\label{sec:d4}

In this Section, we focus on the theory  $(\widehat D_4, 1)$ and explain how the general ansatz (\ref{eq:swcdmansatz}) can be further restricted to obtain the unique Seiberg-Witten curve. As we shall explain, our results agree with  \cite{Haghighat:2018dwe,Haghighat:2018gqf}, which acts as a verification of the general ansatz presented in the previous Subsection~\ref{subsec:genansatz}. We shall furthermore discuss modular properties of our solution and connect the result to known $5$d SCFTs.  

\subsection{Construction}\label{subsec:constructiond4}
We start from the general ansatz~\eqref{eq:swcdmansatz} for $M=4$
{\allowdisplaybreaks
\begin{align}
    &\theta_1^2(z_2;\rho) \left[ a_{00}^{(+)} X_0^2 + a_{11}^{(+)} X_1^2 + a_{02}^{(+)} X_0X_2 + a_{11}^{(-)} Y_1^2 +a_{22}^{(+)} X_2^2 + a_{01}^{(+)} X_0 X_1 + a_{12}^{(+)}X_1 X_2\right]\nonumber\\
    & + \theta_4^2(z_2;\rho) \left[ b_{00}^{(+)} X_0^2 + b_{11}^{(+)} X_1^2 + b_{02}^{(+)} X_0X_2 + b_{11}^{(-)} Y_1^2 +b_{22}^{(+)} X_2^2 + b_{01}^{(+)} X_0 X_1 + b_{12}^{(+)}X_1 X_2 \right]=0\,.\label{D4concrete}
\end{align}}
In order to rearrange this expression into the form~\eqref{eq:TMPHI}, we systematically use Riemann's addition formula (\ref{eq:RiemannP}) in Appendix~\ref{Sect:ThetaFunctions} for genus $1$ theta functions \cite{birkenhake2004complex}. This allows us to systematically replace the basis functions $X_iX_j$ and $Y_iY_j$ by the degree 8 theta functions $\Phi_k$ (defined in~\eqref{eq:defPHI}), as is explicitly shown in eq.~\eqref{eq:XXYYPHI} of Appendix~\ref{Sect:ThetaFunctions}. Indeed, using the shorthand notation:
\begin{equation}
    \phi_i := \Phi_i(0,\tau),\quad \Phi_i := \Phi_i(z_1,\tau), \quad \theta_1:=\theta_1(z_2;\rho),\quad \theta_4:=\theta_4(z_2;\rho),
\end{equation}
we find that (\ref{D4concrete}) can equivalently be written as
\begin{equation}\label{eq:D4ansatz}
\begin{split}
    &\left[(a_{00}^{(+)}\theta_1^2 + b_{00}^{(+)}\theta_4^2)\phi_0 + \left((a_{11}^{(+)}-a_{11}^{(-)})\theta_1^2 + (b_{11}^{(+)}-b_{11}^{(-)})\theta_4^2\right)\frac{\phi_2}{2} + (a_{22}^{(+)}\theta_1^2 + b_{22}^{(+)}\theta_4^2)\phi_4 \right]\Phi_0\\
    &+\left[(a_{01}^{(+)}\theta_1^2 + b_{01}^{(+)}\theta_4^2) \phi_1 + (a_{12}^{(+)}\theta_1^2 + b_{12}^{(+)}\theta_4^2) \phi_3 \right] \Phi_1 \\
    & +\left[\left((a_{11}^{(+)}+a_{11}^{(-)})\theta_1^2 + (b_{11}^{(+)}+b_{11}^{(-)})\theta_4^2\right)\frac{\phi_0+\phi_4}{2} + (a_{02}^{(+)} \theta_1^2 + b_{02}^{(+)}\theta_4^2)\phi_2\right]\Phi_2 \\
    &+\left[(a_{01}^{(+)}\theta_1^2 + b_{01}^{(+)}\theta_4^2) \phi_3 + (a_{12}^{(+)}\theta_1^2 + b_{12}^{(+)}\theta_4^2) \phi_1 \right] \Phi_3\\
    &+\left[(a_{00}^{(+)}\theta_1^2 + b_{00}^{(+)}\theta_4^2)\phi_4 + \left((a_{11}^{(+)}-a_{11}^{(-)})\theta_1^2 + (b_{11}^{(+)}-b_{11}^{(-)})\theta_4^2\right)\frac{\phi_2}{2} + (a_{22}^{(+)}\theta_1^2 + b_{22}^{(+)}\theta_4^2)\phi_0 \right]\Phi_4 = 0\,.
\end{split}
\end{equation}
 In order to further write~\eqref{eq:D4ansatz} in the form of~\eqref{eq:TMPHI}, we need to identify the coefficients of $\Phi_{i}$ (for $i\in\{0,\ldots,4\}$) with quantities $ \sum_{0\leq i,j\leq M} \widetilde T_i(z_2,\rho,\tau)\, \mathbb{M}_{ij}(\tau)$, where in particular the $z_2$ dependent part needs to respect the correct assignment of $SU(2)$ and $SU(1)$ characters: by identifying (some of) the coefficients $a_{ij}^{\pm}$ and $b_{ij}^{(\pm)}$ we have to create combinations of the $\theta_{1,4}^2$, which can be identified with the characters (\ref{eq:defTi}). A priori, there are multiple possibilities to do this, however, not all of them lead to a consistent structure (\ref{eq:TMPHI}) with one $SU(2)$ and 4 $SU(1)$ characters. However, the number of possibilities can be further reduced by assuming additional symmetries:\footnote{In fact, by fixing some of the $a_{ij}^{\pm}$ and $b_{ij}^{(\pm)}$ we have found $7$ independent candidates of the form $\alpha\theta_1^2 + \beta \theta_4^2$ for the $5$ $\widetilde T_i$ entering~\eqref{eq:TMPHI}. We have carefully analysed all of these possible solutions and found that the only consistent ones are compatible with this assumption. Furthermore, we shall motivate this assumption by modular considerations in the following Subsection~\ref{subsec:modularD4}.} we assume that the modular matrix $\mathbb M$ defined by~\eqref{eq:TMPHI} remains invariant under the exchanges 
 \begin{align}
& \widetilde T_0 \leftrightarrow \widetilde T_4,\hspace{0.2cm} \Phi_0 \leftrightarrow \Phi_4\,,&&\text{and} &&\widetilde T_1 \leftrightarrow \widetilde T_3,\hspace{0.2cm} \Phi_1 \leftrightarrow \Phi_3\,.\label{SymExchange}
\end{align}
This is equivalent to stating that none of the Jacobi theta functions multiplied by $a_{ij}^{\pm}$ or $b_{ij}^{\pm}$ with $(ij)\in\{(00)\,, (22)\,, (01)\,, (12)\}$ can be formed into $SU(2)$ characters such that
\begin{equation}
    b_{00}^{(+)} = b_{22}^{(+)} = b_{01}^{(+)} = b_{12}^{(+)} = 0\,,
\end{equation}
and (\ref{eq:D4ansatz}) takes the form
\begin{align}
    &\left[a_{00}^{(+)}\theta_1^2\,\phi_0 + \left(\text{{\color{red}{$(a_{11}^{(+)}-a_{11}^{(-)})\theta_1^2 + (b_{11}^{(+)}-b_{11}^{(-)})\theta_4^2$}}}\right)\frac{\phi_2}{2} + a_{22}^{(+)} \theta_1^2\,\phi_4 \right]\Phi_0+\left[a_{01}^{(+)} \, \phi_1 + a_{12}^{(+)}\,\phi_3 \right]\,\theta_1^2\, \Phi_1\nonumber\\
    & +\left[\left(\text{{\color{blue}{$(a_{11}^{(+)}+a_{11}^{(-)})\theta_1^2 + (b_{11}^{(+)}+b_{11}^{(-)})\theta_4^2$}}}\right)\frac{\phi_0+\phi_4}{2} + \left(\text{{\color{green!50!black}{$a_{02}^{(+)} \theta_1^2 + b_{02}^{(+)}\theta_4^2$}}}\right)\phi_2\right]\Phi_2 +\left[a_{01}^{(+)}\,  \phi_3 + a_{12}^{(+)} \, \phi_1 \right]\theta_1^2 \Phi_3\nonumber\\
    & +\left[a_{00}^{(+)}\theta_1^2 \,\phi_4 + \left(\text{{\color{red}{$(a_{11}^{(+)}-a_{11}^{(-)})\theta_1^2 + (b_{11}^{(+)}-b_{11}^{(-)})\theta_4^2$}}}\right)\frac{\phi_2}{2} + a_{22}^{(+)}\theta_1^2 \,\phi_0 \right]\Phi_4 = 0\,.\label{SWC4Colour}
\end{align}
Among the terms highlighted in colour, one (combination) needs to assume the role of the $SU(2)$ character. This can be achieved in two inequivalent and consistent ways

\begin{enumerate}[label=\Roman*]
    \item\!\!\!) Setting $a_{11}^{(+)}=b_{11}^{(+)}=a_{02}^{(+)}=b_{02}^{(+)}=0$ such that the red and blue terms in (\ref{SWC4Colour}) are identified and play the role of the $SU(2)$ character, while the green term is eliminated:
\begin{align}
    &\left[a_{00}^{(+)}\theta_1^2\,\phi_0 - \left(a_{11}^{(-)}\theta_1^2 + b_{11}^{(-)}\theta_4^2\right)\frac{\phi_2}{2} + a_{22}^{(+)} \theta_1^2\,\phi_4 \right]\Phi_0+\left[a_{01}^{(+)} \, \phi_1 + a_{12}^{(+)}\,\phi_3 \right]\theta_1^2 \Phi_1\nonumber\\
    & +\left[\left(a_{11}^{(-)}\theta_1^2 + b_{11}^{(-)}\theta_4^2\right)\frac{\phi_0+\phi_4}{2} \right]\Phi_2 +\left[a_{01}^{(+)}\,  \phi_3 + a_{12}^{(+)} \, \phi_1 \right]\theta_1^2 \Phi_3 \nonumber\\
    &+\left[a_{00}^{(+)}\theta_1^2 \,\phi_4 - \left(a_{11}^{(-)}\theta_1^2 + b_{11}^{(-)}\theta_4^2\right)\frac{\phi_2}{2} + a_{22}^{(+)}\theta_1^2 \,\phi_0 \right]\Phi_4 = 0\,.\label{SolForm1}
\end{align}  
The same form can also be achieved by identifying $a_{02}^{(+)} = -a_{11}^{(+)}\frac{\phi_0 + \phi_4}{\phi_2}$ and $b_{02}^{(+)} = -b_{11}^{(+)}\frac{\phi_0 + \phi_4}{\phi_2}$). The curve (\ref{SolForm1}) can indeed be cast into the form (\ref{eq:TMPHI}) with
\begin{equation} \widetilde{\mathbf T}^T = \begin{bmatrix}
        a_{00}^{(+)} \theta_1^2 \\ a_{01}^{(+)} \theta_1^2 \\ \tfrac{1}{2}(a_{11}^{(-)} \theta_1^2 + b_{11}^{(-)} \theta_4^2) \\ a_{12}^{(+)} \theta_1^2 \\ a_{22}^{(+)} \theta_1^2
    \end{bmatrix},\quad \mathbb M_{\widehat{D}_4,\rm I}(\tau) =
    \begin{bmatrix}\label{eq:MID4}
        \phi_0 & 0 & 0 & 0 & \phi_4\\
        0 & \phi_1 & 0 & \phi_3 & 0 \\
        -\phi_2 & 0 & \phi_0 + \phi_4 & 0 & -\phi_2\\
        0 & \phi_3 & 0 & \phi_1 & 0 \\
        \phi_4 & 0 & 0& 0 & \phi_0\\
    \end{bmatrix}.
    \end{equation}
In fact, this curve is the result obtained in \cite{Haghighat:2018dwe,Haghighat:2018gqf} since the terms proportional to $\theta_4^2$ are $\beta_2\left(\phi_2\Phi_0 - (\phi_0+\phi_4)\Phi_2 + \phi_2 \Phi_4\right) = 2\beta_2 Y_1^2$ which is indeed compatible with~\cite[eq.~(3.86)]{Haghighat:2018gqf}. 
    \item\!\!\!) Setting $a_{02}^{(+)} = b_{02}^{(+)} = 0$ and identifying $a_{11}^{(-)} = a_{11}^{(+)}$ and $b_{11}^{(-)} = b_{11}^{(+)}$, such that the red and green terms are eliminated and the curve becomes
\begin{align}
    &\left[a_{00}^{(+)}\,\phi_0 + a_{22}^{(+)} \,\phi_4 \right]\theta_1^2\,\Phi_0+\left[a_{01}^{(+)} \, \phi_1 + a_{12}^{(+)}\,\phi_3 \right]\,\theta_1^2\, \Phi_1 +\left[a_{11}^{(+)}\theta_1^2 + b_{11}^{(+)}\theta_4^2\right] (\phi_0+\phi_4)\, \Phi_2 \nonumber\\
    &+\left[a_{01}^{(+)}\,  \phi_3 + a_{12}^{(+)} \, \phi_1 \right]\theta_1^2 \Phi_3 +\left[a_{00}^{(+)} \,\phi_4  + a_{22}^{(+)} \,\phi_0 \right]\,\theta_1^2\,\Phi_4 = 0\,.
\end{align}    
This curve can be cast into the form (\ref{eq:TMPHI}) with
    \begin{equation}\label{eq:MIID4}
    \widetilde{\mathbf T}^T = \begin{bmatrix}
        a_{00}^{(+)} \theta_1^2 \\ a_{01}^{(+)} \theta_1^2 \\ a_{11}^{(+)} \theta_1^2 + b_{11}^{(+)} \theta_4^2 \\ a_{12}^{(+)} \theta_1^2 \\ a_{22}^{(+)} \theta_1^2
    \end{bmatrix},\quad \mathbb M_{\widehat{D}_4,\rm II}(\tau) =
    \begin{bmatrix}
        \phi_0 & 0 & 0 & 0 & \phi_4\\
        0 & \phi_1 & 0 & \phi_3 & 0 \\
        0 & 0 & \phi_0 + \phi_4 & 0 & 0\\
        0 & \phi_3 & 0 & \phi_1 & 0 \\
        \phi_4 & 0 & 0 & 0 & \phi_0\\
    \end{bmatrix}.
    \end{equation}
\end{enumerate}
So far, both solutions are compatible with all symmetries of the $(\widehat{D}_4,1)$ theory and cannot be transformed into one another through any obvious reparametrisations or symmetries. We shall see, however, in Subsection~\ref{subsec:D4lowd} that case I) is compatible with known results of the 5 dimensional theory and thus is the correct SWC to describe the $(\widehat D_4,1)$ LST. Concerning case II), it is not clear whether it describes a consistent LST (or any 6 dimensional gauge theory).

\subsection{Modular properties}\label{subsec:modularD4}
Before discussing the 5 dimensional limit of the $(\widehat{D}_4,1)$ LST, we first discuss modular properties of the two potential SWC characterised by (\ref{eq:MID4}) and (\ref{eq:MIID4}). Since modular transformations acting on the parameter $\rho$ (and coordinate $z_2$) only act on the elliptic Weyl characters $\widetilde{T}_i$ (and are thus relatively straight-forward), we shall focus on modular transformations acting on $\tau$.

From the perspective of the dual theory (\emph{i.e.} the right quiver in Figure~\ref{fig:DMAN}), the gauge polynomials \eqref{eq:sso2m} transform covariantly under an $SL(2,\mathbb{Z})$ symmetry acting on the modular parameter $\tau$ (and the coordinate $z_1$), owing to the transformations \eqref{eq:TStheta1} of the Jacobi theta functions. However, the picture is more complicated from the perspective of the ($\widehat{D}_4,1)$ LST form in eq.~(\ref{eq:TMPHI}), where in fact only a subgroup of this $SL(2,\mathbb{Z})$ is manifestly realised. In the following we shall therefore discuss only transformations under the congruence subgroup $\Gamma_0(4)$,\footnote{This group also corresponds to the modular subgroup acting on the coefficients in the change of basis between Weierstrass functions and theta functions given in~\cite{Nekrasov:2012xe}.} which is the largest subgroup we have found that preserves the general structure (\ref{eq:TMPHI}) of the SWC. $\Gamma_0(4)$ is generated by
\begin{align}
&T = \begin{bmatrix}
        1 & 1 \\ 0 & 1
    \end{bmatrix}\,,&& C= \begin{bmatrix}
        -1 & 0 \\ 0 & -1
    \end{bmatrix}\,,&&\widetilde S_4 = C\cdot S \cdot T^4 \cdot S= \begin{bmatrix}
        1 & 0 \\ -4 & 1
    \end{bmatrix}\,,
\end{align}
where the action of $S$ and $T$ on Jacobi theta functions is explained in Appendix~\ref{app:modulargroup} and the set $\mathbf{\Phi}:=\{\Phi_1,\ldots,\Phi_4\}^T$ is invariant under $C$ by construction (which effectively acts as $z_1\to-z_1$). We start by discussing the action of $T$ on $\mathbb M$ and $\mathbf{\Phi}$. We can treat \eqref{eq:MID4} and~\eqref{eq:MIID4} simultaneously by introducing
\begin{align}
&\epsilon=\left\{\begin{array}{ll}-1 & \text{for I)}\,,\\[4pt] 0& \text{for II)}\,, \end{array}\right.&&(\alpha_2,\beta_2)=\left\{\begin{array}{ll}\left(\tfrac{a_{11}^{(-)}}{2},\tfrac{b_{11}^{(-)}}{2}\right) & \text{for I)}\,,\\[8pt] \left(a_{11}^{(+)},b_{11}^{(+)}\right)& \text{for II)}\,, \end{array}\right.&&\mathbb M_{\widehat D_4}(\epsilon)=\left[\begin{smallmatrix}
        \phi_0 & 0 & 0 & 0 & \phi_4\\
        0 & \phi_1 & 0 & \phi_3 & 0 \\
        \epsilon\phi_2 & 0 & \phi_0 + \phi_4 & 0 & \epsilon\phi_2\\
        0 & \phi_3 & 0 & \phi_1 & 0 \\
        \phi_4 & 0 & 0& 0 & \phi_0\\
    \end{smallmatrix} \right]\,,
\end{align}
and $(\alpha_0,\alpha_1,\alpha_3,\alpha_4)=\left(a_{00}^{(+)},a_{01}^{(+)}, a_{12}^{(+)},a_{22}^{(+)}\right)$, such that the SWC can be written in the form
\begin{equation}
    \theta_1^2(z_2;\rho) \begin{bmatrix}\alpha_0&\alpha_1&\alpha_2&\alpha_3&\alpha_4\end{bmatrix}\cdot \mathbb M_{\widehat D_4}(\epsilon) \cdot {\bf \Phi} + \theta_4^2(z_2;\rho) \beta_2 (\epsilon \phi_2 (\Phi_0 + \Phi_4) + (\phi_0 + \phi_4)\Phi_2 )=0\,,\label{UnifiedSWC4}
\end{equation}
covering both cases. Indeed, under the action of $T$, the matrix $\mathbb{M}_{\widehat{D}_4}(\epsilon)$ transform as
\begin{equation}\label{eq:D4Gamma4T}
     \mathbb M_{\widehat D_4}(\epsilon)\underset{T}{\longrightarrow} \begin{bmatrix}
        \phi_0 & 0 & 0 & 0 & \phi_4\\
        0 & e^{i\pi/4}\phi_1 & 0 & e^{i\pi/4}\phi_3 & 0 \\
        \epsilon i\phi_2 & 0 & i(\phi_0 + \phi_4) & 0 & \epsilon i\phi_2\\
        0 & -e^{i\pi/4}\phi_3 & 0 & -e^{i\pi/4}\phi_1 & 0 \\
        \phi_4 & 0 & 0& 0 & \phi_0
    \end{bmatrix},
\end{equation}
which can be compensated by the transformations
\begin{align}
&\alpha_2\to -i\alpha_2\,, &&\beta_2\to-i\beta_2\,, &&\alpha_1 \to e^{-i\pi/4}\alpha_1\,,&&\alpha_3 \to -e^{-i\pi/4} \alpha_3\,,.
\end{align}
Therefore $T$ leaves the overall form of the SWC invariant for both cases I) and II).

The generator $\widetilde S_4$ acts in the following way on $\mathbb M_{\widehat D_4}(\epsilon)$ 
\begin{equation}\label{eq:D4Gamma4S4}
    \mathbb M_{\widehat D_4}(\epsilon) \underset{\widetilde S_4}{\longrightarrow} \begin{bmatrix}
        \phi_4 & 0 & 0 & 0 & \phi_0\\
        0 & \phi_3 & 0 & \phi_1 & 0 \\
        \epsilon\phi_2 & 0 & \phi_0 + \phi_4 & 0 & \epsilon\phi_2\\
        0 & \phi_1 & 0 & \phi_3 & 0 \\
        \phi_0 & 0 & 0& 0 & \phi_4
    \end{bmatrix}\,,
\end{equation}
which can be compensated by the permutations
\begin{align}
&\alpha_0 \longleftrightarrow \alpha_4\,,&&\text{and}&& \alpha_1 \longleftrightarrow \alpha_3\,.
\end{align}
Therefore also $\widetilde{S}_4$ leaves the overall form of the SWC invariant for both cases I) and II), which are therefore both manifestly invariant under $\Gamma_0(4)$.

We remark, that invariance under $\widetilde{S}_4$ (as explained above) only applies to matrices $\mathbb{M}$ that remain invariant under the exchanges (\ref{SymExchange}). This property therefore serves as additional motivation for imposing this condition when searching for viable forms of the SWC in the previous Subsection.

\subsection{Lower Dimensional Limit}\label{subsec:D4lowd}
So far, both solutions I) and II) have been compatible with all symmetries of the $(\widehat{D}_{4},1)$ LST and are therefore completely equivalent. However, we shall now demonstrate that only solution I) correctly reproduces known results in the literature upon decompactification to 5 dimensions. 

To describe the 5 dimensional limit, we shall use the same notation as in the previous Subsection to treat the two cases I) and II) in parallel. 
First, we take the limit $Q_\rho= e^{2\pi i\rho} \to 0$ (in which the $(\widehat{D}_{4},1)$ LST is described in terms of a conformal field theory):  let 
\begin{align}
&w=e^{2\pi iz_1}\,,&&\text{and}&&t= e^{2\pi iz_2}\,,
\end{align}
and assume that\footnote{Here we assume a scaling of all parameters of the theory, which preserves as much structure as possible of the 6 dimensional theory. Indeed, a weaker scaling of $\beta_2$ would imply that the lower dimensional theory has a different gauge group structure.} $\beta_2 = -\Qr^{-1/4}\widetilde{\beta_2}+\mathfrak{o}(\Qr^{-1/4})$ (while $\alpha_i=\widetilde{\alpha}_i+\mathfrak{o}(\Qr^0)$), then the leading order contribution of (\ref{UnifiedSWC4}) takes the form
\begin{equation}
    (t -1)^2 \begin{bmatrix}\widetilde{\alpha}_0&\widetilde{\alpha}_1&\widetilde{\alpha}_2&\widetilde{\alpha}_3&\widetilde{\alpha}_4\end{bmatrix}\cdot \mathbb M_{\widehat D_4} \cdot {\bf \Phi} + t \widetilde{\beta}_2 (\epsilon \phi_2 (\Phi_0 + \Phi_4) + (\phi_0 + \phi_4)\Phi_2 )=0\,.
\end{equation}
We next extract the leading order in the limit $\Qt =e^{2\pi i\tau}\to 0$, which we shall perform in two steps: first we expand the $\{\Phi_i\}_{i\in\{0\dots M\}}$ in powers of $\Qt$
\begin{align}
0&=4 \widetilde{\alpha}_0 \left[1+\mathcal{O}(\Qt)\right]+ \widetilde{\alpha}_1 \left[2(w + w^{-1}) \Qt^{\frac{1}{8}}+\mathcal{O}(\Qt^{9/8})\right] + \widetilde{\alpha}_2 \left[2(w^2 +2\epsilon + w^{-2}) \Qt^{\frac{1}{4}}+\mathcal{O}(\Qt^{5/4})\right]\nonumber\\
    &\hspace{0.5cm}+ \widetilde{\alpha}_3 \left[2(w^3 + w^{-3}) \Qt^{\frac{5}{8}}+\mathcal{O}(\Qt^{13/8})\right] + \widetilde{\alpha}_4 \left[4(w^4 + w^{-4})\Qt+\mathcal{O}(\Qt^2)\right]\nonumber\\
    &\hspace{0.5cm}+\widetilde{\beta}_2 \left[2(w^2 +2\epsilon + w^{-2}) \Qt^{\frac{1}{4}}+\mathcal{O}(\Qt^{5/4})\right]\,.\label{ScalingBehaviour}
\end{align}
In order to perform the full limit, we need to make certain assumptions regarding the $\Qt$ dependence of the parameters $\widetilde{\alpha}_{0,\ldots,4}$ and $\widetilde{\beta}_2$. As in the case of the $\Qr\to 0$ limit, we shall assume a behaviour that preserves as much of the 6 dimensional symmetries as possible. Concretely, we maintain the exchange symmetries
\begin{align}
&\widetilde{\alpha}_0\longleftrightarrow\widetilde{\alpha}_4\,,&&\text{and}&&\widetilde{\alpha}_1\longleftrightarrow\widetilde{\alpha}_3\,,
\end{align}
which enforces that $\widetilde{\alpha}_0$ and $\widetilde{\alpha}_4$ (as well as $\widetilde{\alpha}_1$ and $\widetilde{\alpha}_3$) have the same leading behaviour in $\Qt$. However, this implies that the contribution with $\widetilde{\alpha}_4$ in (\ref{ScalingBehaviour}) (as well as the contribution with $\widetilde{\alpha}_3$) is subleading and therefore drops out in the limit $\Qt\to 0$. In order to ensure that none of the remaining terms drop out in the limit (which we would interpret as a signal of a modification of the symmetries in the lower dimensional theory), we assume the following scaling behaviour
\begin{align}
&\widetilde{\alpha}_0=\widetilde{\alpha}'_0\,\Qt^{s}+\mathfrak{o}(\Qt^s)\,,&&\widetilde{\alpha}_1=\widetilde{\alpha}'_1\,\Qt^{s-\frac{1}{8}}+\mathfrak{o}(\Qt^{s-1/8})\,,\nonumber\\
&\widetilde{\alpha}_2=\widetilde{\alpha}'_2\,\Qt^{s-\frac{1}{4}}+\mathfrak{o}(\Qt^{s-1/4})\,,&&\widetilde{\beta}_2=\widetilde{\beta}'_2\,\Qt^{s-\frac{1}{4}}+\mathfrak{o}(\Qt^{s-1/4})\,.
\end{align}
where $s\in\mathbb{R}$ is undetermined and shall not be important in the following. We thus obtain the following limit for the SWC
\begin{equation}\label{eq:5dswcd4}
    (t-1)^2 \left(w^2 + w^{-2} + \frac{\widetilde{\alpha}'_1}{\widetilde{\alpha}'_2} (w + w^{-1}) + 2\frac{\widetilde{\alpha}'_0 + \epsilon \widetilde{\alpha}'_2}{\widetilde{\alpha}'_2} \right) + t \frac{\widetilde{\beta}'_2}{\widetilde{\alpha}'_2} \left( w^2 +2\epsilon +w^{-2} \right) = 0.
\end{equation}
This curve can be directly compared with the SWC of the $Sp(N)$ with $N_f$ fundamentals discussed in \cite{Hayashi:2023boy}: let
\begin{align}
&M_1 + M_1^{-1} + M_2 + M_2^{-1} = -\frac{\widetilde{\alpha}'_1}{\widetilde{\alpha}'_2}\,,&& (M_1 + M_1^{-1})(M_2 + M_2^{-1}) +2 = \frac{2\widetilde{\alpha}'_0}{\widetilde{\alpha}'_2}+2\epsilon,
\end{align}
such that we can rewrite (\ref{eq:5dswcd4}) in the language of~\cite{Hayashi:2023boy}:
\begin{align}
    t^2 p_2(w) + \Bigg(\frac{\widetilde{\beta}'_2}{\widetilde{\alpha}'_2}&(w^2 + 2\epsilon + w^{-2})+ 2 \left(M_1 + M_1^{-1} + M_2 + M_2^{-1}\right)(w +w^{-1})\nonumber\\
    &- 2(M_1 + M_1^{-1})(M_2 + M_2^{-1}) -4 -2 (w^2+w^{-2}) \Bigg)t + p_2(w^{-1}) = 0\,,\label{eq:5dswcd4v2}
\end{align}
where we have defined
\begin{equation}
    p_2(w) := w^{-2} \prod_{i=1}^2 (w - M_i)(w-M_i^{-1})=p_2(w^{-1})\,.
\end{equation}
In order to match the contribution of order $t$ in~\eqref{eq:5dswcd4v2} (with $(1-M_i^\pm) = (1-M_i)(1-M_i^{-1})$)
\begin{equation}
    p_1(w) = -\frac{(w+1)^2\prod_{i=1}^2 (1-M_i^\pm)}{2w} +\frac{(w-1)^2\prod_{i=1}^2 (1+M_i^\pm)}{2w} + \frac{\beta_2}{\alpha_2}(w^2 +2\epsilon + w^{-2}) -2(w^2-2+w^{-2}),
\end{equation}
to \cite{Hayashi:2023boy} requires $\epsilon=-1$ and $\beta_2/\alpha_2 = 2 + q^{-1}$, where $q$ is the exponentiated complex gauge coupling of $Sp(0)$. In this way, we  can indeed interpret the resulting SWC as that of a $Sp(0)$ with $SO(4)$ flavor symmetry $5$d SCFT~\cite{Hayashi:2015vhy,Hayashi:2017btw} since $p_2(w)$ has the structure of the corresponding $SO(4)$ gauge polynomial. This is indeed a consistent lower-dimensional limit of the 6 dimensional $(\widehat{D}_4,1)$ LST. Since this fixes $\epsilon=-1$, this also suggests that the correct SWC of the latter theory is in fact described by case I) of Subsection~\ref{subsec:constructiond4}, \emph{i.e.} eq.~(\ref{eq:MID4}). This is also compatible with the result obtained in \cite{Haghighat:2018dwe,Haghighat:2018gqf}.

\section{Seiberg-Witten Curve of $(\widehat D_5,1)$}\label{sec:d5}
After having discussed in detail the construction of the SWC of the $(\widehat D_4,1)$ LST in the previous Section, we now focus on the $(\widehat D_5,1)$ theory, for which the result is not known in the literature. Following the same steps as in the previous case, we show that we find again a unique solution for the SWC.

\subsection{Construction} \label{subsec:consd5}
We explain the construction of the $(\widehat D_5,1)$ LST SWC following the same general ansatz (\ref{eq:swcdmansatz}) outlined in Section~\ref{subsec:genansatz}. Using the change of basis~\eqref{eq:XXYYPHI2}, we can reformulate this curve in the basis $\{\Phi_i\}_{i\in \{0\ldots 5\}}$ of degree 10 theta functions
\begin{equation}\label{eq:d5general}
\begin{split}
    &\left[ ( a_{00}^{(+)}\theta_1^2 +  b_{00}^{(+)}\theta_4^2) \phi_0 + \Xi_{11}^{-}\phi_2 + \Xi_{22}^{-}\phi_4 \right]\Phi_0 + \left[ ( a_{01}^{(+)}\theta_1^2 +  b_{01}^{(+)}\theta_4^2) \phi_1 + \Xi_{12}^{-}\phi_3 + \Xi_{22}^{+}\phi_5 \right]\Phi_1 +\\
    &\left[ ( a_{02}^{(+)}\theta_1^2 +  b_{02}^{(+)}\theta_4^2) \phi_2 + \Xi_{12}^{+}\phi_4 + \Xi_{11}^{+}\phi_5 \right]\Phi_2 +\left[ ( a_{02}^{(+)}\theta_1^2 +  b_{02}^{(+)}\theta_4^2) \phi_3 + \Xi_{12}^{+}\phi_1 + \Xi_{11}^{+}\phi_0 \right]\Phi_3 +\\
    &\left[ ( a_{01}^{(+)}\theta_1^2 +  b_{01}^{(+)}\theta_4^2) \phi_4 + \Xi_{12}^{-}\phi_2 + \Xi_{22}^{+}\phi_0 \right]\Phi_4 +\left[ ( a_{00}^{(+)}\theta_1^2 +  b_{00}^{(+)}\theta_4^2) \phi_5 + \Xi_{11}^{-}\phi_3 + \Xi_{22}^{-}\phi_1 \right]\Phi_5=0\,,
\end{split}
\end{equation}
where we have defined
\begin{equation}
\Xi_{ij}^{\pm} := \frac{a_{ij}^{(+)}\pm a_{ij}^{(-)}}{2}\theta_1^2 +\frac{b_{ij}^{(+)}\pm b_{ij}^{(-)}}{2}\theta_4^2.    
\end{equation}
To simplify the assignment of $SU(2)$ and $SU(1)$ Weyl characters, we shall assume the symmetries\footnote{We motivate this symmetry by the natural $\mathbb Z_2$ action of the $SU(1)$ nodes of the quiver which are labelled by $\widetilde{T}_{0,1,4,5}$.}
\begin{align}
& \widetilde T_i \leftrightarrow \widetilde T_j,\hspace{0.2cm} \Phi_i \leftrightarrow \Phi_j\,, &&\text{for} &&(i,j)\in\{(0,5)\,,(1,4)\}\,,
\end{align}
analogous to (\ref{SymExchange}) in the case of the $(\widehat{D}_4,1)$ LST. These symmetries impose
\begin{equation}
    b_{00}^{(+)}=b_{01}^{(+)}=a_{02}^{(+)}=b_{02}^{(+)}=0.
\end{equation}
and we are left with three candidates ($\Xi_{11}^{\pm},\Xi_{12}^{\pm}$ and $\Xi_{22}^{\pm}$) for two $SU(2)$ characters and two $SU(1)$ characters. The natural choice that provides a symmetry among the $SU(1)$ nodes is to break one of the $\Xi_{ij}^\pm$ (for $(ij)\in\{(11),(12),(22)\}$) into two $SU(1)$ characters. There is a priori three possibilities, schematically:
\begin{align}
    \Xi_{11}^\pm &\to SU(1)^2 &  \Xi_{11}^\pm &\to SU(2) & \Xi_{11}^\pm &\to SU(2)\nonumber\\
    1)\quad\Xi_{12}^\pm &\to SU(2), &  2)\quad\Xi_{12}^\pm &\to SU(1)^2, & 3)\quad\Xi_{12}^\pm &\to SU(2).\nonumber\\
    \Xi_{22}^\pm &\to SU(2) &  \Xi_{22}^\pm &\to SU(2) & \Xi_{22}^\pm &\to SU(1)^2\label{D5Solutions}
\end{align}
All three choices lead to SWC that are compatible with all symmetries of the $(\widehat{D}_5,1)$ LST and (as we shall discuss in the next Subsection), have interesting modular properties: we shall in particular see that there exist modular transformations that transform the solutions 2) and 3) into one another. Furthermore, we shall see in Subsection~\ref{subsec:D5lowd}, that choice 3) provides a (natural) $\Qr,\Qt \to 0$ limit that is compatible with known 5 dimensional results in the literature. Since this is a strong indication that 3) is the correct 6 dimensional SWC of the $(\widehat{D}_5,1)$ LST, we provide here the explicit details of this solution:
\begin{equation}
    a_{11}^{(+)}=b_{11}^{(+)}=a_{22}^{(+)}=b_{22}^{(+)}=b_{12}^{(+)}=b_{12}^{(-)}=0,
\end{equation}
such that
\begin{equation}\label{eq:D5res}\widetilde{\mathbf T}^T = \begin{bmatrix}
        a_{00}^{(+)} \theta_1^2 \\ (a_{22}^{(+)} + a_{22}^{(-)}) \theta_1^2 \\ a_{11}^{(-)} \theta_1^2 + b_{11}^{(-)} \theta_4^2 \\a_{12}^{(-)} \theta_1^2 + b_{12}^{(-)} \theta_4^2 \\ a_{01}^{(+)} \theta_1^2 \\ (a_{22}^{(+)} - a_{22}^{(-)}) \theta_1^2
    \end{bmatrix},\quad \mathbb M_{\widehat D_5}(\tau) =
    \begin{bmatrix}
         \phi_0 & 0 & 0 & 0 & 0 & \phi_5\\
         0 & \phi_5 & 0 & 0 & \phi_0 & 0\\
         -\phi_2 & 0 & \phi_0 & \phi_5 & 0 & -\phi_3\\
         0 & -\phi_3 & \phi_4 & \phi_1 & -\phi_2 & 0\\
         0 & \phi_1 & 0 & 0 & \phi_4 & 0\\
         \phi_4 & 0 & 0 & 0 & 0 & \phi_1
    \end{bmatrix}.
\end{equation}

\subsection{Modular properties}\label{subsec:modD5}
In the same way as in the $(\widehat{D}_4,1)$ LST in Subsection~\ref{subsec:modularD4}, we indicate the action of the generators of a $\Gamma_0(5)$ modular group (acting on the modular parameter $\tau$ and coordinate $z_1$) on the solutions (\ref{D5Solutions}). The congruence subgroup $\Gamma_0(5)\subset SL(2,\mathbb{Z})$ is generated by:
\begin{equation}
    T= \begin{bmatrix}
        1 & 1 \\ 0 & 1
    \end{bmatrix},\quad \widetilde S_{5,1} = ST^2ST^3ST^2 = \begin{bmatrix}
        -3 & -5 \\ 5 & 8
    \end{bmatrix}, \quad \widetilde S_{5,2} = ST^5S = \begin{bmatrix}
        -1 & 0 \\ 5 & -1
    \end{bmatrix}.
\end{equation}
The overall form of $\mathbb M$ is left untouched by the action of $T$ and $\widetilde S_{5,2}$. However, $\widetilde S_{5,2}\circ\widetilde S_{5,1}$ acts non-trivially on $\bf \Phi$ and $\mathbb M_{\widehat D_5}$ given in~\eqref{eq:D5res} (see~\eqref{eq:TSPHI}), in particular

\begin{equation}
    \begin{bmatrix}
         \phi_0 & 0 & 0 & 0 & 0 & \phi_5\\
         0 & \phi_5 & 0 & 0 & \phi_0 & 0\\
         -\phi_2 & 0 & \phi_0 & \phi_5 & 0 & -\phi_3\\
         0 & -\phi_3 & \phi_4 & \phi_1 & -\phi_2 & 0\\
         0 & \phi_1 & 0 & 0 & \phi_4 & 0\\
         \phi_4 & 0 & 0 & 0 & 0 & \phi_1
    \end{bmatrix} \underset{\widetilde S_{5,2}\circ\widetilde{S}_{5,1}}{\longrightarrow} \begin{bmatrix}
         \phi_0 & 0 & 0 & 0 & 0 & \phi_5\\
         0 & 0 & \phi_0 & \phi_5 & 0 & 0\\
         -\phi_4 & \phi_5 & 0 & 0 & \phi_0 & -\phi_1\\
         0 & \phi_3 & -\phi_4 & -\phi_1 & \phi_2 & 0\\
         0 & 0 & \phi_2 & \phi_3 & 0 & 0\\
         \phi_2 & 0 & 0 & 0 & 0 & \phi_3
    \end{bmatrix}
\end{equation}
This matrix, however, corresponds precisely to the solution called in 2) in (\ref{D5Solutions}), indicating that the two are not independent, but are related through a non-trivial modular transformation (and are thus simply equivalent presentations of the same curve.)

\subsection{Lower dimensional limit}\label{subsec:D5lowd}
In order to compute the 5 dimensional limit of the SWC characterised by (\ref{eq:D5res}), we follow the same double limit $\Qr\to 0$ and $\Qt\to 0$ as explained in Section~\ref{subsec:D4lowd} in the case of the $(\widehat{D}_4,1)$ LST, which preserves as much of the higher dimensional symmetries as possible. In the interest of brevity, we shall only display the final result, which we can cast in the form
\begin{equation}
\begin{split}
    (t-1)^2&\left[(w^3+w^{-3}) + \frac{\alpha_2}{\alpha_3}(w^2 +w^{-2}) + \left(\frac{\alpha_1}{\alpha_3}-1\right)(w+w^{-1}) + 2\frac{\alpha_0}{\alpha_3} - 2\frac{\alpha_2}{\alpha_3} \right]\\
    &\quad \quad +\left[ \frac{\beta_3}{\alpha_3}(w^3 - (w+w^{-1}) + w^{-3}) + \frac{\beta_2}{\alpha_3}(w^2 - 2 + w^{-2})  \right]t = 0,
\end{split}
\end{equation}
In order to compare to the results of \cite{Hayashi:2023boy}, we define
\begin{equation}\label{eq:Malpharel}
\begin{split}
    &\frac{\alpha_2}{\alpha_3} = -(M_1 + M_1^{-1} + M_2 + M_2^{-1} +M_3 + M_3^{-1}),\\
    &\frac{\alpha_1}{\alpha_3} = (1+M_1M_2 + M_1M_3 + M_2M_3)(1+(M_1M_2)^{-1}+(M_1M_3)^{-1}+(M_2M_3)^{-1}),\\
    &\frac{\alpha_0}{\alpha_3} = - \frac{(1+M_1^2)(1+M_2^2)(1+M_3^2)}{2M_1M_2M_3}.
\end{split}
\end{equation}
such that the lower dimensional curve can be written in the form
\begin{equation}
    t^2 p_2(w) +\left[ \frac{\beta_3}{\alpha_3}(w^3 - (w+w^{-1}) + w^{-3}) + \frac{\beta_2}{\alpha_3}(w^2 - 2 + w^{-2}) -2p_2(w)  \right]t + p_2(w) = 0,
\end{equation}
with
\begin{equation}
    p_2(w) = \prod_{i=1}^3(w-(M_i+M_i^{-1}) +w^{-1}).
\end{equation}
Following \cite{Hayashi:2023boy}, we redefine $t\to p_2(w)^{-1} t$, the SWC becomes:
\begin{equation}
\begin{split}
    t^2 + \Big[\left( \frac{\beta_3}{\alpha_3}-2\right)&(w^3+w^{-3}) + \left(\frac{\beta_2-2\alpha_2}{\alpha_3} \right)(w^2+w^{-2}) \\
    &+\left(- \chi_c + 2 -\frac{\beta_3}{\alpha_3}\right)(w+w^{-1})+2\left(\chi_s - \frac{\beta_2-2\alpha_2}{\alpha_3} \right)\Big]t + p_2(w)^2 = 0,
\end{split}
\end{equation}
where we used the definition of $SO(2N_f)$ characters for spinor and conjugate spinor representation \cite{Hayashi:2017btw}:
\begin{subequations}
\begin{equation}
    \chi_s := \frac{1}{2} \left( \prod_{i=1}^{N_f} \left(M_i^{-\frac{1}{2}}+M_i^{\frac{1}{2}} \right) + \prod_{i=1}^{N_f} \left(M_i^{-\frac{1}{2}}-M_i^{\frac{1}{2}} \right) \right),
\end{equation}
\begin{equation}
    \chi_c := \frac{1}{2} \left( \prod_{i=1}^{N_f} \left(M_i^{-\frac{1}{2}}+M_i^{\frac{1}{2}} \right) - \prod_{i=1}^{N_f} \left(M_i^{-\frac{1}{2}}-M_i^{\frac{1}{2}} \right) \right).
\end{equation}
\end{subequations}
We can match the result of \cite[eq.~(4.31)]{Hayashi:2023boy} by identifying $ U_2 = q(\beta_2-2\alpha_2)/\alpha_3$ and $\beta_3/\alpha_3 = 2 + q^{-1}$. Furthermore, we impose $M_3=M_3^{-1}=-1$, which has the interpretation of an exponentiated complex mass parameter, which leads to $\chi_c = -\chi_s$. This is a natural choice since in the $\Qt \to 0$ limit the original $SO(10)$ group is broken down to $SO(5)$ which involves two Coulomb branch moduli, there is therefore a spurious parameter in~\eqref{eq:Malpharel}. Thus we obtain: 
\begin{equation}
    t^2 + q^{-1} \left((w^3 +w^{-3}) + U_2(w^2 +w^{-2})  + (q\chi_s -1)(w+w^{-1}) -2(q\chi_c + U_2)\right)t + p_2(w)^2 = 0.
\end{equation}
This result matches the SWC of the $Sp(1)$ with $SO(5)$ flavor symmetry $5$d SCFT~\cite{Hayashi:2015vhy,Hayashi:2017btw} as a specialisation of $Sp(1)$ with $6$ flavours. This therefore serves as a strong consistency check, that (\ref{eq:D5res}) provides the correct SWC of the $(\widehat{D}_5,1)$ LST.

\section{Generalisation to the $(\widehat{D}_M,1)$ LSTs}\label{sec:generalisation}
In the same manner as for the $(\widehat{D}_4,1)$ and $(\widehat{D}_5,1)$ theories, we can analyse $(\widehat{D}_M,1)$ LSTs for generic $M\geq 4$. Here we only indicate a general pattern we have found by studying explicit cases up to $M=12$: In the following we simply state the form of the modular matrix $\mathbb{M}$ and~$\widetilde{\mathbf{T}}$. 

For $M$ odd, we distinguish the cases
\begin{itemize}
\item $M=4M^\prime +1$ 
\begin{equation}
    \widetilde{\mathbf{T}}^T = \begin{bmatrix}
        \alpha_0 \theta_1^2&
        \alpha_1 \theta_1^2&
        \alpha_2 \theta_1^2 + \beta_2 \theta_4^2&
        \cdots&
        \alpha_{4M^\prime -1} \theta_1^2 + \beta_{4M^\prime -1} \theta_4^2&
        \alpha_{4M^\prime} \theta_1^2&
        \alpha_{4M^\prime+1} \theta_1^2
    \end{bmatrix}\!,
\end{equation}
\begin{equation}
    \mathbb M_{\widehat D_{4M^\prime + 1}} =
    \begin{bsmallmatrix}
        \phi_0 & 0 &0 & \cdots & 0 & 0 & \cdots&0 & 0 & \phi_{4M^\prime+1}\\
        0 & \phi_{4M^\prime+1} &0& \cdots & 0 & 0 & \cdots&0 & \phi_0 & 0\\
        -\phi_2 & 0 & \phi_0& \cdots & 0 &0 & \cdots & \phi_{4M^\prime+3} & 0 & -\phi_{4M^\prime-1}\\
        \vdots & \vdots & \vdots & \ddots  & \vdots & \vdots & \iddots &\vdots& \vdots & \vdots\\
        -\phi_{2M^\prime+1} & 0 & 0& \cdots &\phi_{0} & \phi_{4M^\prime+1} & \cdots&0 & 0 & -\phi_{2M^\prime}\\
        0 & -\phi_{2M^\prime} & 0& \cdots &\phi_{4M^\prime} & \phi_{1} & \cdots&0 & -\phi_{2M^\prime+1} & 0\\
        \vdots & \vdots & \vdots &\iddots & \vdots & \vdots & \ddots&\vdots & \vdots & \vdots\\
        0 & -\phi_{4M^\prime-1} & \phi_{4M^\prime+2} & \cdots & 0&0 & \cdots & \phi_1 & -\phi_2 & 0\\
        0 & \phi_1&0 & \cdots & 0 & 0 &\cdots&0 & \phi_{4M^\prime} & 0\\
        \phi_{4M^\prime}&0 & 0 & \cdots & 0 & 0 &\cdots&0 & 0 & \phi_1
    \end{bsmallmatrix}.
\end{equation}
\item $M=4M^\prime +3$:
\begin{equation}
    \widetilde{\mathbf{T}}^T = \begin{bmatrix}
        \alpha_0 \theta_1^2&
        \alpha_1 \theta_1^2&
        \alpha_2 \theta_1^2 + \beta_2 \theta_4^2&
        \cdots&
        \alpha_{4M^\prime+1} \theta_1^2 + \beta_{4M^\prime+1} \theta_4^2&
        \alpha_{4M^\prime+2} \theta_1^2&
        \alpha_{4M^\prime+3} \theta_1^2
    \end{bmatrix}\!,
\end{equation}
\begin{equation}
    \mathbb M_{\widehat D_{4M^\prime + 3}} =
    \begin{bsmallmatrix}
        \phi_0 & 0 &0 & \cdots & 0 & 0 & \cdots&0 & 0 & \phi_{4M^\prime+3}\\
        0 & \phi_{4M^\prime+3} &0& \cdots & 0 & 0 & \cdots&0 & \phi_0 & 0\\
        -\phi_2 & 0 & \phi_0& \cdots & 0 &0 & \cdots & \phi_{4M^\prime+3} & 0 & -\phi_{4M^\prime+1}\\
        \vdots & \vdots & \vdots & \ddots  & \vdots & \vdots & \iddots &\vdots& \vdots & \vdots\\
        -\phi_{2M^\prime+2} & 0 & 0& \cdots &\phi_{4M^\prime+3} & \phi_0 & \cdots&0 & 0 & -\phi_{2M^\prime+1}\\
        0 & -\phi_{2M^\prime+1} & 0& \cdots &\phi_1 & \phi_{4M^\prime+2} & \cdots&0 & -\phi_{2M^\prime+2} & 0\\
        \vdots & \vdots & \vdots &\iddots & \vdots & \vdots & \ddots&\vdots & \vdots & \vdots\\
        0 & -\phi_{4M^\prime+1} & \phi_{4M^\prime+2} & \cdots & 0&0 & \cdots & \phi_1 & -\phi_2 & 0\\
        0 & \phi_1&0 & \cdots & 0 & 0 &\cdots&0 & \phi_{4M^\prime+2} & 0\\
        \phi_{4M^\prime+2}&0 & 0 & \cdots & 0 & 0 &\cdots&0 & 0 & \phi_1
    \end{bsmallmatrix}.
\end{equation}
\end{itemize}
while for $M$ even we distinguish
\begin{itemize}
\item $M=4M^\prime$ 
\begin{equation}
    \widetilde{\mathbf{T}} = \begin{bmatrix}
        \alpha_0 \theta_1^2&
        \alpha_1 \theta_1^2&
        \alpha_2 \theta_1^2 + \beta_2 \theta_4^2&
        \cdots&
        \alpha_{4M^\prime -2} \theta_1^2 + \beta_{4M^\prime -2} \theta_4^2&
        \alpha_{4M^\prime-1} \theta_1^2&
        \alpha_{4M^\prime} \theta_1^2
    \end{bmatrix}\!,
\end{equation}
\begin{equation}
    \mathbb M_{\widehat D_{4M^\prime}} =
    \begin{bsmallmatrix}
        \phi_0 & 0 & 0 & \cdots & 0  & 0 & \cdots & 0 & \phi_{4M^\prime}\\
        0 & \phi_{1} & 0 & \cdots & 0  & \cdots & 0 & \phi_{4M^\prime-1} & 0\\
        -\phi_2 & 0 & \phi_2& \cdots & 0 & \cdots & \phi_{4M^\prime} & 0 & -\phi_{4M^\prime-2}\\
        \vdots & \vdots & \vdots & \ddots  & \vdots & \iddots & \vdots & \vdots & \vdots\\
        -\phi_{2M^\prime} & 0 &0 & \cdots &\phi_0 + \phi_{4M^\prime}  & \cdots & 0 & 0 & -\phi_{2M^\prime}\\
        \vdots & \vdots & \vdots &\iddots & \vdots & \ddots & \vdots & \vdots & \vdots\\
        -\phi_{4M^\prime-2} & 0 & \phi_{4M^\prime} & \cdots & 0 & \cdots & \phi_0 & 0 & -\phi_2\\
        0 & \phi_{4M^\prime-1} & 0 & \cdots & 0  &\cdots & 0 & \phi_1 & 0\\
        \phi_{4M^\prime} & 0 &0 & \cdots & 0  &\cdots & 0 & 0 & \phi_0
    \end{bsmallmatrix},
\end{equation}
\item $M=4M^\prime +2$ 
\begin{equation}
    \widetilde{\mathbf{T}} = \begin{bmatrix}
        \alpha_0 \theta_1^2&
        \alpha_1 \theta_1^2&
        \alpha_2 \theta_1^2 + \beta_2 \theta_4^2&
        \cdots&
        \alpha_{4M^\prime } \theta_1^2 + \beta_{4M^\prime } \theta_4^2&
        \alpha_{4M^\prime+1} \theta_1^2&
        \alpha_{4M^\prime+2} \theta_1^2
    \end{bmatrix}\!,
\end{equation}
\begin{equation}
    \mathbb M_{\widehat D_{4M^\prime + 2}} =
    \begin{bsmallmatrix}
        \phi_0 & 0 & 0 & \cdots & 0  & 0 & \cdots & 0 & \phi_{4M^\prime}\\
        0 & \phi_{1} & 0 & \cdots & 0  & \cdots & 0 & \phi_{4M^\prime-1} & 0\\
        -\phi_2 & 0 & \phi_2& \cdots & 0 & \cdots & \phi_{4M^\prime} & 0 & -\phi_{4M^\prime-2}\\
        \vdots & \vdots & \vdots & \ddots  & \vdots & \iddots & \vdots & \vdots & \vdots\\
        0 & -\phi_{2M^\prime+1} &0 & \cdots &\phi_1 + \phi_{4M^\prime+1}  & \cdots & 0 & -\phi_{2M^\prime+1} & 0\\
        \vdots & \vdots & \vdots &\iddots & \vdots & \ddots & \vdots & \vdots & \vdots\\
        -\phi_{4M^\prime-2} & 0 & \phi_{4M^\prime} & \cdots & 0 & \cdots & \phi_0 & 0 & -\phi_2\\
        0 & \phi_{4M^\prime-1} & 0 & \cdots & 0  &\cdots & 0 & \phi_1 & 0\\
        \phi_{4M^\prime} & 0 &0 & \cdots & 0  &\cdots & 0 & 0 & \phi_0
    \end{bsmallmatrix}.
\end{equation}
\end{itemize}
Explicit examples for the matrices $\mathbb{M}$ for $M\in\{6 \ldots 9\}$ can be found in Appendix~\ref{app:Mmat}.
\paragraph{}
These four classes of result can be expressed is a unifying manner from the dual perspective of $\widehat A_1$ base theory (\emph{i.e.} the right quiver in Figure~\ref{fig:DMAN}). The results translate to the elliptic gauge polynomials of $SO(2M)$ and $Sp(M^\prime)$ with $M^\prime=M-4$ and can be understood as a direct generalisation of the case $M=4$~\eqref{eq:A1SOSP}. Elliptic gauge polynomials are finite products of Jacobi theta functions and can be systematically understood as a Higgsing of a $SU(N)$ elliptic gauge polynomial~\cite{Hollowood:2003cv}:
\begin{align}
    s_{SU(N)}(z_1;\tau) = \prod_{i=1}^N \frac{\theta_1(z_1-a_i;\tau)}{\theta_1(z_1;\tau)},\quad \text{and} \quad \langle \varphi \rangle = \left(a_{1},\ldots, a_{N}\right),
\end{align}
$\langle \varphi \rangle$ corresponds to the gauge holonomies of $SU(N)$. For $SO(2M)$, we obtain:
\begin{equation}
    F_1(z_1;\tau) = s_{SO(2M)}(z_1,\tau) = \prod_{i=1}^M \frac{\theta_1(z_1-a_i^{SO};\tau)\theta_1(z_1+a_i^{SO};\tau)}{\theta_1^2(z_1;\tau)}\,.\label{GaugePotential1}
\end{equation}
which corresponds to a Higgsing of $SU(2M)$ gauge polynomial given by:
\begin{equation}
    \langle \varphi \rangle = \left(\pm a_{1}^{SO},\ldots, \pm a_{M}^{SO} \right).
\end{equation}
For $Sp(M^\prime)$, we have:
\begin{equation}
    F_2(z_1;\tau) = s_{Sp(M^\prime)}(z_1,\tau) = \frac{\theta_1^2(2z_1;\tau)}{\theta_1^8(z_1;\tau)} \prod_{i=1}^{M^\prime} \frac{\theta_1(z_1-a_i^{Sp};\tau)\theta_1(z_1+a_i^{Sp};\tau)}{\theta_1^2(z_1;\tau)}\,,\label{GaugePotential2}
\end{equation}
which corresponds to a Higgsing of $SU(2M^\prime + 8)$ gauge polynomial given by:
\begin{equation}
    \langle \varphi \rangle = \left(0,0,\pm \frac{1}{2},\pm \frac{\tau}{2},\pm \frac{1+\tau}{2},\pm a_{1}^{Sp},\ldots, \pm a_{M^\prime}^{Sp} \right).
\end{equation}
We remark that $F_2$ can be seen as a specialisation of $F_1$ by using~\eqref{eq:theta2zid} and identifying $\left(a_1^{SO},\ldots,a_4^{SO}\right) = \left(0,\frac{1}{2},\frac{\tau}{2},\frac{\tau+1}{2}\right)$ (which correspond to the so-called discrete holonomies). This result is in line with observations in lower dimensions. In $5$ dimensions, $SO$ or $Sp$ gauge theories with flavors can be engineered from Type IIB String Theory through $5$-brane webs and $O7^\pm$ orientifold planes \cite{Hayashi:2023boy}. Typically, $5$d $\mathcal{N}=1$ $SO$ theories are constructed using $O7^+$-planes while $Sp$ theories arise from constructions involving $O7^-$. In \cite{Hayashi:2023boy}, it was pointed out that an $O7^+$ plane admits an equivalent description as an $O7^-$ along with $8 D7\big|_{\rm frozen}$ i.e. with no moduli attached to them. We interpret the relation between $F_1$ and $F_2$ as a $6$d manifestation of this observation.

\section{Conclusion}\label{Sect:Conclusions}
In this paper we have constructed the Seiberg-Witten Curves (SWCs) for a family of Little String Theories (LSTs) engineered by a single M5-brane probing a transverse (binary) dihedral orbifold geometry $\mathbb{R}_\perp^4/\mathbb{BD}_{M-2}$ (with $M\geq 4$). The six-dimensional world-volume theory on the M5-brane can be described by a quiver gauge theory, as shown in the left part of Figure~\ref{fig:DMAN}: the quiver is in the form of the Dynkin diagram of $\widehat{D}_{M}$, with nodes of type $SU(2)$ and $SU(1)$ respectively and matter in the bifundamental representations. This theory, denoted $(\widehat{D}_M,1)$ in this work, was shown to be dual \cite{Kim:2017xan} (see also \cite{Haghighat:2018dwe} for the case $M=4$) to a circular quiver gauge theory (\emph{i.e.} with quiver group $\widehat{A}_0$) with one node of type $Sp(M-4)$ and one node of type $SO(2M)$ and bifundamental matter, as shown in the right part of Figure~\ref{fig:DMAN}. While a form for the SWC of the theory $(\widehat{D}_M,1)$ with $M=4$ was written down in  \cite{Haghighat:2018dwe,Haghighat:2018gqf}, a construction for general $M$ is not known in the literature.

In (\ref{eq:swcdmansatz}) we provide a general ansatz for the SWC for generic $M\geq 4$ that is compatible with important symmetries and dualities, notably a decomposition in the form of equation~(\ref{eq:TMPHI}) in terms of (elliptic) Weyl characters $\widetilde{\mathbf{T}}$ associated with the different gauge nodes, a basis $\{\Phi_j\}_{j\in\{0,\ldots,M\}}$ of $\mathbb{Z}_2$-even theta functions of degree $2M$ (see~\eqref{eq:defPHI}) and a modular matrix $\mathbb{M}_{ij}$. The ansatz (\ref{eq:swcdmansatz}) depends on more than the $2M-2$ expected parameters, such that further conditions need to be imposed to determine the SWC: indeed, further parameters are fixed by demanding that the curve can also be cast in the form (\ref{eq:A1SOSP}), thus reflecting the dual description of the $(\widehat{D}_{M},1)$ theory in terms of the $Sp(M-4)$--$SO(2M)$ quiver gauge theory explained above. Finally, by assuming a natural dimensional reduction to 5 dimensions, and comparison with known results of the lower dimensional SWC \cite{Hayashi:2023boy} further puts restrictions on the possible solutions. In this way, we have found in Sections~\ref{sec:d4} and \ref{sec:d5} respectively, a unique solution for the SWC of the $(\widehat{D}_4,1)$ and $(\widehat{D}_5,1)$ respectively. In the former case, the solution precisely agrees with the curve presented in \cite{Haghighat:2018dwe,Haghighat:2018gqf}, while the latter constitutes a novel result. 

We have systematically analysed in a similar fashion all $(\widehat{D}_{M},1)$ LSTs up to $M=12$, which present a number of clear structured patterns (see Appendix~\ref{app:Mmat} for the modular matrices up to $M=10$). These have lead us to conjecture the general form of the $(\widehat{D}_{M},1)$ SWC, as explained in Section~\ref{sec:generalisation}. Furthermore, using the dual form of the SWC, we also find the form (\ref{GaugePotential1}) and (\ref{GaugePotential2}) respectively, of the gauge potentials $F_{1,2}$ entering into (\ref{eq:A1SOSP}).

By determining the form of the SWC of the $(\widehat{D}_{M},1)$ theory for generic $M$, our work casts further light on the $\widehat{D}$-type LSTs, which are far less studied than their $\widehat{A}$-type counterparts \cite{Hollowood:2003cv,Braden:2003gv}. For example, the relation between the two (equivalent) presentations (\ref{eq:TMPHI}) and (\ref{eq:A1SOSP}) encodes details on the duality between the $(\widehat{D}_{M},1)$ and the $Sp(M-4)$--$SO(2M)$ theory.

Our results may also hint to further $\widehat{D}$-type orbifold theories in 6 dimensions. Indeed, in the case of $M=4$ (Section~\ref{sec:d4}) and $M=5$ (Section~\ref{sec:d5}), the final form of the SWC could only be determined after taking a (natural) decompactification limit and comparison with lower dimensional results. In both cases, we found alternative forms of the curve ((\ref{eq:MIID4}) and (\ref{D5Solutions}) respectively) that are compatible with all symmetries of the problem, and only whose natural dimensional reduction leads to a different curve than expected in the lower-dimensional theory \cite{Hayashi:2023boy}. It is an interesting question, whether these curves nevertheless correspond to viable SWC of -- possibly deformed -- theories in 6 dimensions and whether they are dual to the $(\widehat{D}_{M},1)$ LSTs. If this is the case, there may exist more intricate limits than the ones considered in Sections~\ref{subsec:D4lowd} and~\ref{subsec:D5lowd}, respectively, that makes them compatible with the known lower dimensional results. We have seen one example of such a scenario being realised, namely the case entitled 2) in (\ref{D5Solutions}), which we showed to be related to the case 3) by a modular transformation (an element of the congruence subgroup $\Gamma_0(5)$). Since the latter SWC in turn reduces by a natural limit to a known 5 dimensional SWC, this implies that there exists also a (transformed) limit, that reduces the case 2) to a viable 5 dimensional result. Similar mechanisms can a priori not be ruled out for all other 6 dimensional SWCs that we found to be compatible with all higher dimensional symmetries, thus leaving the question whether they indeed realise viable gauge theories. We leave this question for future research.

Let us remark another potential application of our results.
It has been established that the Coulomb branch of the supersymmetric vacua of eight supercharge theory is identified with the phase space of the algebraic integrable system~\cite{Donagi:1995cf,Seiberg:1996nz}. 
In this context, the SWC is identified with the spectral curve of the corresponding integrable system.
Since the integrable system associated with the $\widehat{A}$-type LST is known to be the double elliptic system~\cite{Braden:1999aj,Mironov:1999vj,Braden:2001yc}, the SWC discussed in this paper would be identified with a spectral curve of new $\widehat D$-type double elliptic integrable systems.

A further interesting question concerns the generalisation of our work to more general LSTs. Indeed, here we have only considered theories engineered by a single M5 brane probing a dihedral orbifold geometry. It would be interesting to study the generalisation to multiple M5-branes, \emph{i.e.} LSTs of the type $(\widehat{D}_M,N)$ for $N\geq 1$. For $N>1$, the existence of a duality generalising Figure~\ref{fig:DMAN} is not evident and it will be interesting to analyse the structure of the SWC in this case. Finally, instead of considering a transverse dihedral orbifold, it would be interesting to consider other transverse group actions, which lead to the $\widehat{E}$-type LSTs.

\section*{Acknowledgements}
We are very grateful to B. Haghighat for useful correspondence on the SWC of the $(\widehat{D}_4,1)$ LST. BF and SH are grateful to the Quantum Theory Center ($\hbar$QTC, D-IAS, University of Southern Denmark) for kind hospitality when this work was being finalised. BF and TK also acknowledge the Galileo Galilei Institute for Theoretical Physics (GGI) for support and hospitality. The work of TK was supported by CNRS through MITI interdisciplinary programs, EIPHI Graduate School (No.~ANR-17-EURE-0002) and Bourgogne-Franche-Comté region.

\appendix

\section{Modular functions and properties}\label{App:ModularPropertie}
In this Appendix, we define modular functions and give some of their properties that are useful in the computation of SWC. 
\subsection{Definitions and properties}\label{Sect:ThetaFunctions}
We use the notation $\Theta$ for genus $2$ theta functions and $\theta$ for genus $1$ theta functions. Genus $2$ theta functions are defined as:
\begin{equation}\label{eq:THETA}
    \Theta \! \begin{bmatrix} { \delta} \\ { \epsilon}\end{bmatrix} \! (  z    |\Omega) = \sum_{ m \in \mathbb Z^2} \exp \left( i \pi ( m +  \delta)^T \cdot \Omega \cdot ( m +  \delta) + 2i \pi ( z +   \epsilon)^T (  m +   \delta) \right),
\end{equation}
and verify:
\begin{equation}\label{eq:GAactiongenus2}
    \begin{split}
        \Theta \! \begin{bmatrix} {\delta}   \\ \mathbf 0  \end{bmatrix} \! (  z    |A\cdot \Omega \cdot A^T) = \Theta \! \begin{bmatrix} {A \cdot \delta}   \\ \mathbf 0  \end{bmatrix} \! ( z  \cdot A^{-1}  |\Omega)\,.
    \end{split}
\end{equation}
We define the standard Jacobi theta function as:
\begin{align}\label{eq:defthetajac}
    &\theta_1(z;\tau) = \theta \! \begin{bmatrix}1/2\\1/2\end{bmatrix}\!(z|\tau), &\theta_2(z;\tau) = \theta \! \begin{bmatrix}1/2\\0\end{bmatrix}\!(z|\tau),\nonumber\\
    &\theta_3(z;\tau) = \theta \! \begin{bmatrix}0\\0\end{bmatrix}\!(z|\tau), &\theta_4(z;\tau) = \theta \! \begin{bmatrix}0\\1/2\end{bmatrix}\!(z|\tau),
\end{align}
that are related by shifts in $\mathbb Z/2 \oplus \tau \mathbb Z/2$:
\begin{align}
    \theta_2(z;\tau) = \theta_1\left(z + \frac{1}{2};\tau \right), && \theta_3(z;\tau) = \theta_4\left(z + \frac{1}{2};\tau \right), &&\theta_1(z;\tau) = -i e^{i\pi z} e^{i\pi \frac{\tau}{4}} \theta_4\left(z+\frac{\tau}{2};\tau\right),
\end{align}
and verify:
\begin{equation}\label{eq:theta2zid}
    \theta_1(z;\tau)\theta_2(z;\tau)\theta_3(z;\tau)\theta_4(z;\tau) = \frac{\theta_1(2z;\tau)}{2\theta_2(0;\tau)\theta_3(0;\tau)\theta_4(0;\tau)}\,.
\end{equation}
In this work, we are interested in basis of modular forms that arise as sections of a degree $r$ line bundle. Elements of such basis $\{\varphi_i(z;\tau)\}_{i\in I}$ will be denoted as degree $r$ modular forms and have the property that each $\varphi_i(z;\tau)/\theta_1(z;\tau)^r$ has a pole of maximal order $r$ in $z$ in the $\tau \to 0$ limit, for all $i \in I$. For example, we define $X_i$ $(Y_i)$, the elements of $\mathbb Z_2$ even (odd) basis of degree $M$ modular forms:
\begin{equation}\label{eq:defxiyi}
\begin{split}
    &X_i(z_1,\tau) = \theta \! \begin{bmatrix}\frac{i}{M}\\0\end{bmatrix}\!(Mz_1|M\tau) + \theta \! \begin{bmatrix}\frac{M-i}{M}\\0\end{bmatrix}\!(Mz_1|M\tau)\,,\quad i \in \{0\ldots \lfloor M/2 \rfloor\}\,,\\
    &Y_i(z_1,\tau) = \theta \! \begin{bmatrix}\frac{i}{M}\\0\end{bmatrix}\!(Mz_1|M\tau) - \theta \! \begin{bmatrix}\frac{M-i}{M}\\0\end{bmatrix}\!(Mz_1|M\tau)\,,\quad i \in \{1\ldots \lfloor (M-1)/2 \rfloor\}\,,
\end{split}
\end{equation}
and $\Phi_i$ the elements of $\mathbb Z_2$ even basis of degree $2M$ modular forms:
\begin{equation}\label{eq:defPHI}
    \Phi_j(z_1,\tau) = \theta \! \begin{bmatrix}\frac{j}{2M}\\0\end{bmatrix}\!(2Mz_1|2M\tau) + \theta \! \begin{bmatrix}\frac{2M-j}{2M}\\0\end{bmatrix}\!(2Mz_1|2M\tau)\,, \quad j \in \{0\ldots M\}\,,
\end{equation}
the two basis are related by Riemann's addition formula
\begin{equation}\label{eq:RiemannP}
    \theta \! \begin{bmatrix}\frac{i}{M}\\0\end{bmatrix}\!(z_1|\tau)\,\theta \! \begin{bmatrix}\frac{j}{M}\\0\end{bmatrix}\!(z_1|\tau) = \sum_{k\in \{0,\frac{1}{2}\}} \theta \! \begin{bmatrix}k+\frac{i-j}{2M}\\0\end{bmatrix}\!(0|2\tau)\,\theta \! \begin{bmatrix}k+\frac{i+j}{2M}\\0\end{bmatrix}\!(2z_1|2\tau)\,.
\end{equation}
In particular, we have the following product rules for $\widehat D_4$:
\begin{align}\label{eq:XXYYPHI}
        &X_0^2 = \phi_0 \Phi_0 + \phi_4 \Phi_4, & &X_2^2 = \phi_0 \Phi_4 + \phi_4 \Phi_0,\nonumber\\
        & X_0 X_2 = 2 \phi_2 \Phi_2, & &X_1 X_2 = \phi_1\Phi_3 + \phi_3\Phi_1,\nonumber\\
        &X_1^2 = \frac{\phi_0+\phi_4}{2} \Phi_2 + \frac{\phi_2}{2}(\Phi_0 + \Phi_4), & &Y_1^2 = \frac{\phi_0+\phi_4}{2} \Phi_2 - \frac{\phi_2}{2}(\Phi_0 + \Phi_4),\nonumber\\
        &X_0X_1 = \phi_1\Phi_1 + \phi_3\Phi_3. &&
\end{align}
and product rules for $\widehat D_5$:
\begin{align}\label{eq:XXYYPHI2}
    &X_0^2 = \phi_0\Phi_0 + \phi_5 \Phi_5,&&X_0 X_1 = \phi_1 \Phi_1 + \phi_4\Phi_4,\nonumber\\
    &X_1^2 = \frac{\phi_0}{2}\Phi_2 + \frac{\phi_5}{2} \Phi_3 + \frac{\phi_4}{2} \Phi_5 + \frac{\phi_1}{2}\Phi_0,&&Y_1^2 = \frac{\phi_0}{2}\Phi_2 + \frac{\phi_5}{2} \Phi_3 - \frac{\phi_4}{2} \Phi_5 - \frac{\phi_1}{2}\Phi_0,\nonumber\\
    &X_1X_2 = \frac{\phi_1}{2}\Phi_3+ \frac{\phi_4}{2} \Phi_2 + \frac{\phi_2}{2}\Phi_4 + \frac{\phi_3}{2}\Phi_1,&&Y_1Y_2 = \frac{\phi_1}{2}\Phi_3+ \frac{\phi_4}{2} \Phi_2 - \frac{\phi_2}{2}\Phi_4 - \frac{\phi_3}{2}\Phi_1 ,\nonumber\\
    &X_2^2 = \frac{\phi_0}{2}\Phi_4 + \frac{\phi_5}{2} \Phi_1 + \frac{\phi_1}{2} \Phi_5 + \frac{\phi_4}{2} \Phi_0,&&Y_2^2 = \frac{\phi_0}{2}\Phi_4 + \frac{\phi_5}{2} \Phi_1 - \frac{\phi_1}{2} \Phi_5 - \frac{\phi_4}{2} \Phi_0,\nonumber\\
    &X_0 X_2 = \phi_2 \Phi_2 + \phi_3 \Phi_3.&&
\end{align}

\subsection{Modular (sub)group}\label{app:modulargroup}
Defining the action of elements of a modular subgroup $\Gamma \subseteq SL(2,\mathbb Z)$ on $(z,\tau)$:
\begin{equation}
    \begin{bmatrix}
        a & b\\ c& d
    \end{bmatrix}\in \Gamma :(z,\tau) \longmapsto \left(\frac{z}{c\tau +d}, \frac{a\tau +b}{c\tau +d} \right),
\end{equation}
the set of theta functions used to formulate the SWC~\eqref{eq:TMPHI} is vector-valued under such modular transformations \cite{eichler1985theory}:
\begin{subequations}
\begin{equation}\label{eq:Ttrans}
    \theta \! \begin{bmatrix}
        \frac{j}{2M}\\0
    \end{bmatrix} \! (2Mz_1|2M(\tau+1)) \underset{T:\tau\to\tau+1}{\longrightarrow} e^{i\pi \frac{j^2}{2M}}\theta \! \begin{bmatrix}
        \frac{j}{2M}\\0
    \end{bmatrix} \! (2Mz_1|2M\tau),
\end{equation}
\begin{equation}\label{eq:Strans}
    \theta \! \begin{bmatrix}
        \frac{j}{2M}\\0
    \end{bmatrix} \! (2Mz_1|2M\tau) \underset{S:z_1,\tau\to\frac{z_1}{\tau},\frac{-1}{\tau}}{\longrightarrow} \omega(z_1|\tau)\sum_{k=0}^{2M-1} e^{-i\pi \frac{kj}{M}}\theta \! \begin{bmatrix}
        \frac{k}{2M}\\0
    \end{bmatrix} \! (2Mz_1|2M\tau),
\end{equation}
\end{subequations}
with $\omega(z_1|\tau) = \sqrt{\frac{\tau}{2iM}} \exp\left(\frac{2iM z_1^2}{\tau}\right)$, $S=\begin{bmatrix}
    0 & -1 \\ 1 & 0
\end{bmatrix}$ and $T = \begin{bmatrix}
    1 & 1 \\ 0 & 1
\end{bmatrix}$.
This generates the following transformation on the base of even modular forms of degree $2M$:
\begin{equation}\label{eq:TSPHI}
    \mathbf{\Phi}:=
    \begin{bmatrix}
        \Phi_0\\\vdots\\\Phi_{M}
    \end{bmatrix} \underset{T:\tau\to\tau+1}{\longrightarrow} \left[ \exp \left(i\pi \frac{j^2}{2M} \right) \right]_{jj} \cdot \mathbf{\Phi}, \quad \mathbf{\Phi} \underset{S:z_1,\tau\to\frac{z_1}{\tau},\frac{-1}{\tau}}{\longrightarrow} \frac{\omega(z_1|\tau)}{\sqrt{2M}}\left[ 2 \cos \left( \frac{ij \pi}{M}\right) \right]_{ij} \cdot {\bf \Phi}.
\end{equation}
Standard Jacobi theta functions are re-scaled by modular transformations, in particular $\theta_1$ verifies:
\begin{equation}\label{eq:TStheta1}
    \theta_1(z_1;\tau) \underset{T:\tau\to\tau+1}{\longrightarrow} e^{i\frac{\pi}{4}}\theta_1(z_1;\tau),\quad \theta_1(z_1;\tau)\underset{S:z_1,\tau\to\frac{z_1}{\tau},\frac{-1}{\tau}}{\longrightarrow} -i \sqrt{\frac{\tau}{i}} e^{i\frac{\pi z_1^2}{\tau}} \theta_1(z_1;\tau).
\end{equation}
We define the modular congruence subgroup $\Gamma_0(M)$ as:
\begin{equation}\label{eq:defgamma0}
    \Gamma_0(M) = \left\{ \begin{bmatrix} a & b \\ c & d \end{bmatrix} \in SL(2,\mathbb Z), \,\, c \equiv 0 \,\, \text{mod}\,\, M \right\}.
\end{equation}

\section{$\mathbb M$ matrices examples}\label{app:Mmat}
In this appendix, we give explicit examples of $\mathbb M$ matrices entering~\eqref{eq:TMPHI} of the SWC for $\widehat D_{6\ldots9}$. For $(\widehat D_6,1)$ LST, we have:
\begin{equation}
    \mathbb{M}_{\widehat D_{6}}(\tau) = \begin{bmatrix}
            \phi_0 & 0 & 0 & 0 & 0 & 0 & \phi_6\\
            0 & \phi_1 & 0 & 0 & 0 & \phi_5 & 0\\
            -\phi_2 & 0 & \phi_0 & 0 & \phi_6 & 0 & -\phi_4\\
            0& -\phi_3 & 0 & \phi_1 + \phi_5 & 0 & -\phi_3 & 0\\
            -\phi_4 & 0 & \phi_6 & 0 & \phi_0 & 0 & -\phi_2\\
            0 & \phi_5 & 0 & 0 & 0 & \phi_1 & 0\\
            \phi_6 & 0 & 0 & 0 & 0 & 0 & \phi_0
        \end{bmatrix},
\end{equation}
for $(\widehat D_7,1)$ LST:
\begin{equation}
    \mathbb M_{\widehat D_7}(\tau) = \begin{bmatrix}
         \phi_0 & 0 & 0 & 0 & 0 & 0 & 0 & \phi_7\\
         0 & \phi_7 & 0 & 0 & 0 & 0 & \phi_0 & 0\\
         -\phi_2 & 0 & \phi_0 & 0 & 0 & \phi_7 & 0 & -\phi_5\\
         -\phi_4 & 0 & 0 & \phi_7 & \phi_0 & 0 & 0 & -\phi_3\\
         0 & -\phi_3 & 0 & \phi_1 & \phi_6 & 0 & -\phi_4 & 0\\
         0 & -\phi_5 & \phi_6 & 0 & 0 & \phi_1 & -\phi_2 & 0\\
         0 & \phi_1 & 0 & 0 & 0 & 0 & \phi_6 & 0\\
         \phi_6 & 0 & 0 & 0 & 0 & 0 & 0 & \phi_1
    \end{bmatrix},
\end{equation}
for $(\widehat D_8,1)$ LST:
\begin{equation}
    \mathbb M_{\widehat D_8}(\tau) = \begin{bmatrix}
            \phi_0 & 0 & 0 & 0 & 0 & 0 & 0 & 0 & \phi_8\\
            0 & \phi_1 & 0 & 0 & 0 & 0 & 0 & \phi_7 & 0\\
            -\phi_2 & 0 & \phi_0 & 0 & 0 & 0 & \phi_8 & 0 & -\phi_6\\
            0& -\phi_3 & 0 & \phi_1 & 0 & \phi_7 & 0 & -\phi_5 & 0\\
            -\phi_4 & 0 & 0 & 0 & \phi_0 + \phi_8 & 0& 0 & 0 & -\phi_4\\
            0 & -\phi_5 & 0 & \phi_7 & 0 & \phi_1 &0 & -\phi_3 & 0\\
            -\phi_6 & 0 & \phi_8 & 0 & 0 & 0 & \phi_0 & 0 & -\phi_2\\
            0 & \phi_7 & 0 & 0 & 0 & 0 & 0 & \phi_1 & 0\\
            \phi_8 & 0 & 0 & 0 & 0 & 0 & 0 & 0 & \phi_0
        \end{bmatrix},
\end{equation}
for $(\widehat D_9,1)$ LST:
\begin{equation}
    \mathbb M_{\widehat D_9}(\tau) = \begin{bmatrix}
         \phi_0 & 0 & 0 & 0 & 0 & 0 & 0 & 0 & 0 & \phi_9\\
         0 & \phi_9 & 0 & 0 & 0 & 0 & 0 & 0 & \phi_0 & 0\\
         -\phi_2 & 0 & \phi_0 & 0 & 0 & 0 & 0 & \phi_9 & 0 & -\phi_7\\
         -\phi_4 & 0 & 0 & \phi_9 & 0 & 0 & \phi_0 & 0 & 0 & -\phi_5\\
         -\phi_6 & 0 & 0 & 0 & \phi_0 & \phi_9 & 0 & 0 & 0 & -\phi_3\\
         0 & -\phi_3 & 0 & 0 & \phi_8 & \phi_1& 0 & 0 & -\phi_6 & 0\\
         0 & -\phi_5 & 0 & \phi_1 & 0 & 0& \phi_8 & 0 & -\phi_4 & 0\\
         0 & -\phi_7 & \phi_8 & 0 & 0 & 0 & 0 & \phi_1 & -\phi_2 & 0\\
         0 & \phi_1 & 0 & 0 & 0 & 0 & 0 & 0 & \phi_8 & 0\\
         \phi_8 & 0 & 0 & 0 & 0 & 0 & 0 & 0 & 0 & \phi_1
    \end{bmatrix}.
\end{equation}

\begingroup
\sloppy
\printbibliography
\endgroup

\end{document}